\documentclass[3p,12pt]{elsarticle}

\usepackage{amssymb,latexsym,amsmath,multirow,url}

\journal{}

\newcommand{\eps}{\varepsilon}

\newcommand{\set}[1]{\left\{#1\right\}}

\newcommand{\p}{\partial}

\newtheorem{thm}{Theorem}[section]

\newtheorem{lem}[thm]{Lemma}

\newproof{pf}{Proof}

\begin{document}

\begin{frontmatter}



\title{Multi-frequency topological derivative for approximate shape acquisition of curve-like thin electromagnetic inhomogeneities}


\author{Won-Kwang Park}
\ead{parkwk@kookmin.ac.kr}
\address{Department of Mathematics, Kookmin University, Seoul, 136-702, Korea.}

\begin{abstract}
In this paper, we investigate a non-iterative imaging algorithm based on the topological derivative in order to retrieve the shape of penetrable electromagnetic inclusions when their dielectric permittivity and/or magnetic permeability differ from those in the embedding (homogeneous) space. The main objective is the imaging of crack-like thin inclusions, but the algorithm can be applied to arbitrarily shaped inclusions. For this purpose, we apply multiple time-harmonic frequencies and normalize the topological derivative imaging function by its maximum value. In order to verify its validity, we apply it for the imaging of two-dimensional crack-like thin electromagnetic inhomogeneities completely hidden in a homogeneous material. Corresponding numerical simulations with noisy data are performed for showing the efficacy of the proposed algorithm.
\end{abstract}

\begin{keyword}
Thin electromagnetic inclusions \sep Topological derivative \sep Multiple frequencies \sep Numerical experiments


\end{keyword}

\end{frontmatter}





\section{Introduction and preliminaries}\label{Sec1}
The main objective of this paper is the development of a topological derivative based one-step iterative imaging algorithm for thin electromagnetic inclusions completely embedded in a homogeneous domain, via boundary measurement. For proper beginning, we review related mathematical models, and corresponding formulas, followed by a brief condensation of recent results and an outline of the current paper.

\begin{figure}
\begin{center}
\includegraphics[width=0.45\textwidth,keepaspectratio=true,angle=0]{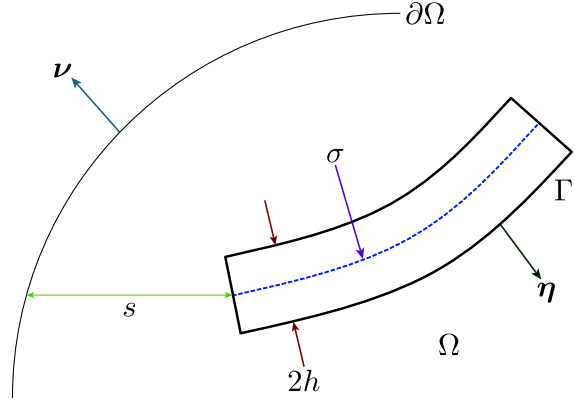}
\caption{\label{FigureGamma}Two-dimensional thin electromagnetic inclusion $\Gamma$ of thickness $2h$.}
\end{center}
\end{figure}

Let $\Omega\subset\mathbb{R}^2$ be a homogeneous domain with smooth boundary $\p\Omega$, which is a $\mathcal{C}^3$ curve; this domain contains a thin, curve-like homogeneous electromagnetic inclusion. Let us assume that this thin inclusion (denoted as $\Gamma$) is represented in the neighborhood of a simple smooth curve $\sigma:=\sigma(\mathbf{x})$ as
\[\Gamma=\set{\mathbf{x}+\gamma\mathbf{n}(\mathbf{x}):\mathbf{x}\in\sigma,~\gamma\in(-h,h)},\]
where $\mathbf{n}(\mathbf{x})$ is the unit normal to $\sigma$ at $\mathbf{x}$ and $h$ is a positive constant that denotes the thickness of $\Gamma$ refer to Figure \ref{FigureGamma}. Throughout this paper, we assume that the applied frequency is of the form $\omega=\frac{2\pi}{\lambda}$ for the given wavelength $\lambda$, the thickness $h$ of $\Gamma$ is sufficiently small with respect to $\lambda$ ($h\ll\lambda$), and the inclusion does not touch the boundary $\p\Omega$ so that it must be located at some distance from $\p\Omega$. In other words, there is a nonzero positive constant $s$ such that
\[\mbox{dist}(\sigma,\p\Omega)=s\gg h.\]

Let every material be classified by its dielectric permittivity and magnetic permeability at a given frequency $\omega$. Let $0<\eps_0<+\infty$ and $0<\mu_0<+\infty$ denote the permittivity and permeability of the domain $\Omega$, and $0<\eps<+\infty$ and $0<\mu<+\infty$, those of the inclusion $\Gamma$. Then, we can define the piecewise constant dielectric permittivity $\varepsilon(\mathbf{x})$ and magnetic permeability $\mu(\mathbf{x})$ as
\begin{equation}\label{EPSMU}
\varepsilon(\mathbf{x})=\left\{\begin{array}{ccl}
\varepsilon_0&\mbox{for}&\mathbf{x}\in\Omega\backslash\overline{\Gamma}\\
\varepsilon&\mbox{for}&\mathbf{x}\in\Gamma
\end{array}\right.
\quad\mbox{and}\quad
\mu(\mathbf{x})=\left\{\begin{array}{ccl}
\mu_0&\mbox{for}&\mathbf{x}\in\Omega\backslash\overline{\Gamma}\\
\mu&\mbox{for}&\mathbf{x}\in\Gamma,
\end{array}\right.
\end{equation}
respectively. For the sake of simplicity, we set $\eps_0=\mu_0=1$, $\eps>\eps_0$, and $\mu>\mu_0$.

At a given frequency $\omega$, let $u^{(l)}(\mathbf{x};\omega)$ be the time-harmonic total field satisfying the Helmholtz equation in the existence of $\Gamma$,
\begin{equation}\label{ForwardProblem}
\left\{\begin{array}{rcl}
\displaystyle\nabla\cdot\left(\frac{1}{\mu(\mathbf{x})}\nabla u^{(l)}(\mathbf{x};\omega)\right)+\omega^2\eps(\mathbf{x})u^{(l)}(\mathbf{x};\omega)=0&\mbox{in}&\Omega\\
\noalign{\medskip}\displaystyle\frac{1}{\mu_0}\frac{\p u^{(l)}(\mathbf{x};\omega)}{\p\boldsymbol{\nu}(\mathbf{x})}=\frac{\p e^{i\omega\mathbf{d}_l\cdot\mathbf{x}}}{\p\boldsymbol{\nu}(\mathbf{x})}=g^{(l)}(\mathbf{x};\omega)\in L^2(\partial\Omega)&\mbox{on}&\p\Omega,\\
\end{array}\right.
\end{equation}
with transmission conditions
\[u^{(l)}(\mathbf{x};\omega)|_+=u^{(l)}(\mathbf{x};\omega)|_-\quad\mbox{and}\quad
\frac{1}{\mu_0}\frac{\p u^{(l)}(\mathbf{x};\omega)}{\p \boldsymbol{\eta}(\mathbf{x})}\bigg|_+=\frac{1}{\mu}\frac{\p u^{(l)}(\mathbf{x};\omega)}{\p \boldsymbol{\eta}(\mathbf{x})}\bigg|_-\quad\mbox{on}\quad\p\Gamma.\]
Here, $\boldsymbol{\nu}(\mathbf{x})$ and $\boldsymbol{\eta}(\mathbf{x})$ represent the unit outward normal to $\mathbf{x}\in\p\Omega$ and $\mathbf{x}\in\p\Gamma$, respectively, subscript $\pm$ denotes the limiting values as
\[u(\mathbf{x})|_\pm=\lim_{t\to0\pm}u(\mathbf{x}\pm t\boldsymbol{\eta}(\mathbf{x}))\quad\mbox{and}\quad
\frac{\p u^{(l)}(\mathbf{x})}{\p \boldsymbol{\eta}(\mathbf{x})}\bigg|_\pm=\lim_{t\to0\pm}\frac{\p u^{(l)}(\mathbf{x}\pm\boldsymbol{\eta}(\mathbf{x}))}{\p \boldsymbol{\eta}(\mathbf{x})}\quad\mbox{for}\quad\mathbf{x}\in\p\Gamma,\]
and $\mathbf{d}_l=(\cos\theta_l,\sin\theta_l)$ denotes a two-dimensional vector on the unit circle $\mathbb{S}^1$. Similarly, let $u_{\mbox{\tiny bac}}^{(l)}(\mathbf{x};\omega)=e^{i\omega\mathbf{d}_l\cdot\mathbf{x}}$ denote a field satisfying (\ref{ForwardProblem}) without $\Gamma$, i.e., a background solution. Throughout this paper, we assume that $\omega^2$ is not an eigenvalue of (\ref{ForwardProblem}).

As mentioned earlier in this section, the main purpose of this paper is to develop a fast, non-iterative algorithm for imaging a thin inclusion $\Gamma$ completely embedded in a domain $\Omega$, via the boundary measurements $u^{(l)}(\mathbf{x};\omega)$, $\mathbf{x}\in\p\Omega$. Note that there is a remarkable number of interesting inverse scattering problems for reconstructing thin electromagnetic inclusions and/or perfectly conducting cracks hidden in a structure (such as bridges, concrete walls, and machine constructions) from boundary measurements, refer to \cite{A1,A2} and references therein. For this purpose, various iterative and non-iterative imaging algorithms have been developed and successfully applied to various problems, for example, level-set method \cite{ADIM,DL,PL4}, MUltiple SIgnal Classification (MUSIC)-type \cite{AGKPS,AK,AKLP,PL1,PL3}, linear sampling method \cite{C,KR} and multi-frequency based algorithms \cite{AGKPS,HSSZ,P1,P2,PL2}. From many researches, it turns out that non-iterative imaging algorithms are fast, simple, effective, and extendable to multiple targets; however, they require a large number of incident directions and boundary measurements. In contrast to the non-iterative algorithms, iterative imaging algorithms do not require a large number of incident directions and boundary measurements. However, they require complex calculation of the so called Fr{\'e}chet derivative, adequate regularization terms for each iteration step, a good initial guess whose shape is close to the unknown target (here, $\Gamma$) and \textit{a priori} information of target, e.g., material properties, thickness, location. Owing to these considerations, the realization of a trade-off between non-iterative and iterative imaging algorithms is an interesting research topic.

Topological derivative strategy has been developed for this purpose. Recently, this strategy was successfully applied to the shape optimization and imaging of small and crack-like inhomogeneities, see \cite{AGJK,AKLP,B,CR,EKS,P3,P4,SZ} for instance. For successful application of this strategy theoretically, a large number of incident directions and corresponding scattered fields are required. Unfortunately, for practical application, it is extremely difficult to increase the number of such fields owing to the high configuration costs, unavoidable random noise, and so on.

The above limitation has motivated us to consider an improved topological derivative for imaging thin, extended electromagnetic inclusions. For this purpose, we propose an imaging functional based on the topological derivative at multiple frequencies. We explore some properties and limitations of traditional topological derivative based imaging functional, and we aim to improve them accordingly.

The remainder of this paper is organized as follows. In section \ref{Sec2}, we briefly introduce the topological derivative based imaging functional derived in \cite{P4}. A normalized multi-frequency imaging functional is proposed in section \ref{Sec3}. In section \ref{Sec4}, we present the results of numerical simulations to illustrate the advantages and disadvantages of the proposed imaging algorithm. Finally, we conclude this paper in section \ref{Sec5}.

\section{Review of normalized topological derivative at single frequency}\label{Sec2}
In this section, we shall introduce the basic concept of topological derivative operated at a fixed single frequency. We would like to mention \cite{AGJK,AKLP,B,CR,EKS,P3,P4,SZ} for detailed discussions. Let $u_{\mbox{\tiny tot}}^{(l)}(\mathbf{x};\omega)$ and $u_{\mbox{\tiny bac}}^{(l)}(\mathbf{x};\omega)$ be the total and background solutions of (\ref{ForwardProblem}), respectively. The problem considered herein is the minimization of the following energy functional depending on the solution $u^{(l)}(\mathbf{x};\omega)$:
\begin{equation}\label{Energy}
  \mathbb{E}(\Omega;\omega):=\frac12\sum_{l=1}^{L}\|u_{\mbox{\tiny tot}}^{(l)}(\mathbf{x};\omega)-u_{\mbox{\tiny bac}}^{(l)}(\mathbf{x};\omega)\|_{L^2(\partial\Omega)}^2=\frac12\sum_{l=1}^{L}\int_{\p\Omega}|u_{\mbox{\tiny tot}}^{(l)}(\mathbf{x};\omega)-u_{\mbox{\tiny bac}}^{(l)}(\mathbf{x};\omega)|^2dS(\mathbf{x}).
\end{equation}

Assume that an electromagnetic inclusion $\Sigma$ of small diameter $r$ is created at a certain position $\mathbf{z}\in\Omega\backslash\partial\Omega$, and let $\Omega|\Sigma$ denote this domain. Since the topology of the entire domain has changed, we can consider the corresponding topological derivative $d_T\mathbb{E}(\mathbf{z})$ based on $\mathbb{E}(\Omega)$ with respect to point $\mathbf{z}$ as
\begin{equation}\label{TopDerivative}
  d_T\mathbb{E}(\mathbf{z};\omega)=\lim_{r\to0+}\frac{\mathbb{E}(\Omega|\Sigma;\omega) -\mathbb{E}(\Omega;\omega)}{\varphi(r;\omega)},
\end{equation}
where $\varphi(r;\omega)\longrightarrow0$ as $r\longrightarrow0+$. From (\ref{TopDerivative}), we can obtain an asymptotic expansion:
\begin{equation}\label{AsymptoticExpansionTopologicalDerivative}
  \mathbb{E}(\Omega|\Sigma;\omega)=\mathbb{E}(\Omega;\omega)+
\varphi(r;\omega)d_T\mathbb{E}(\mathbf{z};\omega)+o(\varphi(r;\omega)).
\end{equation}

In \cite{P4}, the following normalized topological derivative imaging function $\mathbb{E}_{\mbox{\tiny TD}}(\mathbf{z};\omega)$ has been introduced:
\begin{equation}\label{NormalizedTopologicalDerivative}
  \mathbb{E}_{\mbox{\tiny TD}}(\mathbf{z};\omega)=\frac{1}{2}\bigg(\frac{d_T\mathbb{E}_\eps(\mathbf{z};\omega)}{\max[d_T\mathbb{E}_\eps(\mathbf{z};\omega)]} +\frac{d_T\mathbb{E}_\mu(\mathbf{z};\omega)}{\max[d_T\mathbb{E}_\mu(\mathbf{z};\omega)]}\bigg).
\end{equation}
Here, $d_T\mathbb{E}_\eps(\mathbf{z};\omega)$ and $d_T\mathbb{E}_\mu(\mathbf{z};\omega)$ satisfying (\ref{AsymptoticExpansionTopologicalDerivative}) for purely dielectric permittivity contrast ($\eps\ne\eps_0$ and $\mu=\mu_0$) and magnetic permeability contrast ($\eps=\eps_0$ and $\mu\ne\mu_0$) cases, respectively, are explicitly expressed as (see \cite{P4})
  \begin{align}
      d_T\mathbb{E}_\eps(\mathbf{z};\omega)&=\mathfrak{Re}\sum_{l=1}^{L}\bigg(v_{\mbox{\tiny adj}}^{(l)}(\mathbf{z};\omega)\overline{u_{\mbox{\tiny bac}}^{(l)}(\mathbf{z};\omega)}\bigg),\label{TopologicalDerivative1}\\
      d_T\mathbb{E}_\mu(\mathbf{z};\omega)&=\mathfrak{Re}\sum_{l=1}^{L}\bigg(\nabla v_{\mbox{\tiny adj}}^{(l)}(\mathbf{z};\omega)\cdot\overline{\nabla u_{\mbox{\tiny bac}}^{(l)}(\mathbf{z};\omega)}\bigg),\label{TopologicalDerivative2}
  \end{align}
where $v_{\mbox{\tiny adj}}^{(l)}(\mathbf{x};\omega)$ satisfies the adjoint problem
\begin{equation}\label{Adjoint1}
  \left\{\begin{array}{rcl}
    \displaystyle\Delta v_{\mbox{\tiny adj}}^{(l)}(\mathbf{x};\omega)+\omega^2v_{\mbox{\tiny adj}}^{(l)}(\mathbf{x};\omega)=0&\mbox{in}&\Omega\\
    \noalign{\medskip}\displaystyle\frac{\p v_{\mbox{\tiny adj}}^{(l)}(\mathbf{x};\omega)}{\p\boldsymbol{\nu}(\mathbf{x})}=u_{\mbox{\tiny tot}}^{(l)}(\mathbf{x};\omega)-u_{\mbox{\tiny bac}}^{(l)}(\mathbf{x};\omega)&\mbox{on}&\p\Omega.
  \end{array}\right.
\end{equation}

Some remarkable properties of (\ref{TopologicalDerivative1}) and (\ref{TopologicalDerivative2}) for small and extended thin electromagnetic inclusions can be found in \cite{AGJK} and \cite{P4}, respectively.

\section{Introduction to normalized multi-frequency topological derivative: theory and calculation}\label{Sec3}
Topological derivative based imaging algorithm is well known for its fast imaging performance and robustness with respect to random noise (see \cite{AGJK} for instance). However, when measured data is affected by a considerable amount of noise and/or the number of incident directions $L$ is small (see \cite{P4} for the effect of $L$), one cannot obtain a good result. In order to address these issues, we refer to multi-frequency based imaging techniques \cite{AGKPS,G,HSSZ,P1,P2,PL2}, and we consider the following normalized multi-frequency based topological derivative imaging function: for several frequencies $\set{\omega_k:k=1,2,\cdots,K}$, define
\begin{equation}\label{MultiFrequencyTopologicalDerivative}
  \mathbb{E}(\mathbf{z};K):=\frac{1}{K}\sum_{k=1}^{K}\mathbb{E}_{\mbox{\tiny TD}}(\mathbf{z};\omega_k)=
  \frac{1}{2K}\sum_{k=1}^{K}\bigg(\frac{d_T\mathbb{E}_\eps(\mathbf{z};\omega_k)}{\max[d_T\mathbb{E}_\eps(\mathbf{z};\omega_k)]} +\frac{d_T\mathbb{E}_\mu(\mathbf{z};\omega_k)}{\max[d_T\mathbb{E}_\mu(\mathbf{z};\omega_k)]}\bigg),
\end{equation}
where $d_T\mathbb{E}_\eps(\mathbf{z};\omega_k)$ and $d_T\mathbb{E}_\mu(\mathbf{z};\omega_k)$ satisfy (\ref{TopologicalDerivative1}) and (\ref{TopologicalDerivative2}), respectively, for $\omega=\omega_k$, $k=1,2,\cdots,K$.

From now on, we will analyze the properties of (\ref{MultiFrequencyTopologicalDerivative}). For this purpose, we recall the following result from \cite{P4}. Note that only a concise proof of Lemma \ref{lem3} is introduced in \cite{P4}; we have provided a detailed proof of Lemma \ref{lem3} in Appendix \ref{SecB}.

\begin{lem}\label{lem3}
  Let $A\sim B$ imply that there exists a constant $C$ such that $A=BC$, and let $\mathfrak{Re}(f)$ denote the real part of $f$. Then, (\ref{TopologicalDerivative1}) and (\ref{TopologicalDerivative2}) satisfy
  \begin{align*}
    d_T\mathbb{E}_\varepsilon(\mathbf{z};\omega_k)&\sim\mathfrak{Re}\sum_{l=1}^{L}\int_\sigma(\eps-\eps_0) e^{i\omega_k\mathbf{d}_l\cdot(\mathbf{x-z})}d\sigma(\mathbf{x})\\
    d_T\mathbb{E}_\mu(\mathbf{z};\omega_k)&\sim\mathfrak{Re}\sum_{l=1}^{L}\int_\sigma \bigg[2\bigg(\frac{1}{\mu}-\frac{1}{\mu_0}\bigg)\mathbf{d}_l\cdot\mathbf{t}(\mathbf{x})
    +2\bigg(\frac{1}{\mu_0}-\frac{\mu}{\mu_0^2}\bigg)\mathbf{d}_l\cdot\mathbf{n}(\mathbf{x})\bigg] e^{i\omega_k\mathbf{d}_l\cdot(\mathbf{x-z})}d\sigma(\mathbf{x}),
  \end{align*}
  where $\mathbf{t}(\mathbf{x})$ and $\mathbf{n}(\mathbf{x})$ are unit vectors that are respectively tangent and normal to the supporting curve $\sigma$ at $\mathbf{x}$.
\end{lem}

With this, we can obtain the following result.

\begin{thm}\label{TheoremMF1} Assume that the number of incident directions $L(\geq4)$ is small and the applied number of frequencies $F$ is finite ($F<+\infty$); then, (\ref{MultiFrequencyTopologicalDerivative}) becomes
\[\mathbb{E}(\mathbf{z};K)\approx\frac{1}{2}\bigg(\frac{\mathbb{E}_1(\mathbf{z};K)}{\max|\mathbb{E}_1(\mathbf{z};K)|}+\frac{\mathbb{E}_2(\mathbf{z};K)}{\max|\mathbb{E}_2(\mathbf{z};K)|}\bigg),\]
where
\begin{align*}
\mathbb{E}_1(\mathbf{z};K)=\sum_{l=1}^{L}\int_{\sigma}&(\eps-\eps_0)j_0\bigg(\frac{\omega_K-\omega_1}{2}\mathbf{d}_l\cdot(\mathbf{x-z})\bigg) \cos\bigg(\frac{\omega_K+\omega_1}{2}\mathbf{d}_l\cdot(\mathbf{x-z})\bigg)d\sigma(\mathbf{x})\\
\mathbb{E}_2(\mathbf{z};K)=\sum_{l=1}^{L}\int_{\sigma}&\bigg[2\bigg(\frac{1}{\mu}-\frac{1}{\mu_0}\bigg)\mathbf{d}_l\cdot\mathbf{t}(\mathbf{x}) +2\bigg(\frac{1}{\mu_0}-\frac{\mu}{\mu_0^2}\bigg)\mathbf{d}_l\cdot\mathbf{n}(\mathbf{x})\bigg]\\
&\times j_0\bigg(\frac{\omega_K-\omega_1}{2}\mathbf{d}_l\cdot(\mathbf{x-z})\bigg) \cos\bigg(\frac{\omega_K+\omega_1}{2}\mathbf{d}_l\cdot(\mathbf{x-z})\bigg)d\sigma(\mathbf{x}),
\end{align*}
and $j_0(x)$ denotes the spherical Bessel function of order zero,
\[j_0(x)=\frac{\sin x}{x}.\]
\end{thm}
\begin{pf}
First, we consider the term $\mathbb{E}_1(\mathbf{z};K)$ by evaluating
\begin{align}
\begin{aligned}\label{epsJ}
  \sum_{k=1}^{K}d_T\mathbb{E}_\varepsilon(\mathbf{z};\omega_k) &\approx\int_{\omega_1}^{\omega_K}{d_T\mathbb{E}_\varepsilon(\mathbf{z})d\omega}
  \approx\mathfrak{Re}\int_{\omega_1}^{\omega_K}(\eps-\eps_0)\bigg[\sum_{l=1}^{L}\int_\sigma e^{i\omega\mathbf{d}_l\cdot(\mathbf{x-z})}d\sigma(\mathbf{x})\bigg]d\omega\\
  &=\sum_{l=1}^{L}\int_{\sigma}(\eps-\eps_0)\bigg(\mathfrak{Re}\int_{\omega_1}^{\omega_K} e^{i\omega\mathbf{d}_l\cdot(\mathbf{x-z})}d\omega\bigg) d\sigma(\mathbf{x}).
\end{aligned}
\end{align}
Performing an elementary calculus yields
\begin{align*}
  \int_{\omega_1}^{\omega_K} e^{i\omega\mathbf{d}_l\cdot(\mathbf{x-z})}d\omega=
  &\bigg[\frac{e^{i\omega\mathbf{d}_l\cdot(\mathbf{x-z})}}{i\mathbf{d}_l\cdot(\mathbf{x-z})}\bigg]_{\omega_1}^{\omega_K}
  =\frac{1}{i\mathbf{d}_l\cdot(\mathbf{x-z})} \bigg[e^{i\omega_K\mathbf{d}_l\cdot(\mathbf{x-z})}-e^{i\omega_1\mathbf{d}_l\cdot(\mathbf{x-z})}\bigg]\\
  =&\frac{1}{i\mathbf{d}_l\cdot(\mathbf{x-z})}
  \bigg[\cos(\omega_K\mathbf{d}_l\cdot(\mathbf{x-z}))-\cos(\omega_1\mathbf{d}_l\cdot(\mathbf{x-z}))\\
  &+i\sin(\omega_K\mathbf{d}_l\cdot(\mathbf{x-z}))-i\sin(\omega_1\mathbf{d}_l\cdot(\mathbf{x-z}))\bigg]\\
  =&\frac{2}{\mathbf{d}_l\cdot(\mathbf{x-z})}
  \bigg[\cos\bigg(\frac{(\omega_K+\omega_1)}{2}\mathbf{d}_l\cdot(\mathbf{x-z})\bigg)
  \sin\bigg(\frac{(\omega_K-\omega_1)}{2}\mathbf{d}_l\cdot(\mathbf{x-z})\bigg)\\
  &+i\sin\bigg(\frac{(\omega_K+\omega_1)}{2}\mathbf{d}_l\cdot(\mathbf{x-z})\bigg)
  \sin\bigg(\frac{(\omega_K-\omega_1)}{2}\mathbf{d}_l\cdot(\mathbf{x-z})\bigg)\bigg].
\end{align*}
Therefore, by taking the real part of the above formula, (\ref{epsJ}) can be approximated as
\begin{align}
\begin{aligned}\label{Eepsilon}
\sum_{k=1}^{K}d_T\mathbb{E}_\varepsilon(\mathbf{z};\omega_k)&\approx
\sum_{l=1}^{L}\int_{\sigma}(\eps-\eps_0)\frac{\sin(\xi_1\mathbf{d}_l\cdot(\mathbf{x-z}))}{\mathbf{d}_l\cdot(\mathbf{x-z})}
\cos(\xi_2\mathbf{d}_l\cdot(\mathbf{x-z}))d\sigma(\mathbf{x})\\
&=\frac{2}{\xi_1}\sum_{l=1}^{L}\int_{\sigma}(\eps-\eps_0)j_0(\xi_1\mathbf{d}_l\cdot(\mathbf{x-z})) \cos(\xi_2\mathbf{d}_l\cdot(\mathbf{x-z}))d\sigma(\mathbf{x}),
\end{aligned}
\end{align}
where
\[\xi_1:=\frac{\omega_K-\omega_1}{2}\quad\mbox{and}\quad\xi_2:=\frac{\omega_K+\omega_1}{2}.\]
Hence, by taking the maximum value of (\ref{Eepsilon}) and using it for normalization, we can obtain the desired structure of $\mathbb{E}_1(\mathbf{z};K)$.

Next, we consider the term $\mathbb{E}_2(\mathbf{z};K)$. Since $\mathbf{d}_l$, $\mathbf{t}(\mathbf{x})$, and $\mathbf{n}(\mathbf{x})$ do not depend on $\omega_k$, we can similarly obtain the following approximation
\begin{align}
\begin{aligned}\label{Emu}
  &\sum_{k=1}^{K}d_T\mathbb{E}_\mu(\mathbf{z};\omega_k)\\
  &\approx\sum_{l=1}^{L}\int_{\sigma}\bigg[2\bigg(\frac{1}{\mu}-\frac{1}{\mu_0}\bigg)\mathbf{d}_l\cdot\mathbf{t}(\mathbf{x}) +2\bigg(\frac{1}{\mu_0}-\frac{\mu}{\mu_0^2}\bigg)\mathbf{d}_l\cdot\mathbf{n}(\mathbf{x})\bigg]\frac{\sin(\xi_1\mathbf{d}_l\cdot(\mathbf{x-z}))}{\mathbf{d}_l\cdot(\mathbf{x-z})}
\cos(\xi_2\mathbf{d}_l\cdot(\mathbf{x-z}))d\sigma(\mathbf{x})\\
  &=\frac{2}{\xi_1}\sum_{l=1}^{L}\int_{\sigma}\bigg[2\bigg(\frac{1}{\mu}-\frac{1}{\mu_0}\bigg)\mathbf{d}_l\cdot\mathbf{t}(\mathbf{x}) +2\bigg(\frac{1}{\mu_0}-\frac{\mu}{\mu_0^2}\bigg)\mathbf{d}_l\cdot\mathbf{n}(\mathbf{x})\bigg]j_0(\xi_1\mathbf{d}_l\cdot(\mathbf{x-z})) \cos(\xi_2\mathbf{d}_l\cdot(\mathbf{x-z}))d\sigma(\mathbf{x}).
  \end{aligned}
\end{align}
By applying the maximum value of (\ref{Emu}), the structure of $\mathbb{E}_2(\mathbf{z};K)$ can be obtained.
\end{pf}

\begin{thm}\label{TheoremMF2}Assume that the number of incident directions $L$ is sufficiently large and the applied number of frequencies $F$ is finite ($F<+\infty$); then, (\ref{MultiFrequencyTopologicalDerivative}) becomes
\[\mathbb{E}(\mathbf{z};K)\approx\frac{1}{2}\bigg(\frac{\mathbb{E}_3(\mathbf{z};K)}{\max|\mathbb{E}_3(\mathbf{z};K)|}+\frac{\mathbb{E}_4(\mathbf{z};K)}{\max|\mathbb{E}_4(\mathbf{z};K)|}\bigg)\]
with
\begin{align*}
  \mathbb{E}_3(\mathbf{z};K)&=2\pi\int_\sigma(\eps-\eps_0)\bigg(\Lambda(t;\omega_K)-\Lambda(t;\omega_1)\bigg)d\sigma(\mathbf{x})\\
  \mathbb{E}_4(\mathbf{z};K)&=2\pi\int_\sigma\bigg[2\bigg(\frac{1}{\mu}-\frac{1}{\mu_0}\bigg)\mathbf{d}_l\cdot\mathbf{t}(\mathbf{x}) +2\bigg(\frac{1}{\mu_0}-\frac{\mu}{\mu_0^2}\bigg)\mathbf{d}_l\cdot\mathbf{n}(\mathbf{x})\bigg]\bigg(\Lambda(t;\omega_K)-\Lambda(t;\omega_1)\bigg)d\sigma(\mathbf{x}).
\end{align*}
Here, $\Lambda(t;\omega)$ is defined as
\begin{equation}\label{FunctionLambda}
  \Lambda(t;\omega):=\omega J_0(\omega t)+\frac{\omega\pi}{2}\bigg(J_1(\omega t)H_0(\omega t)-J_0(\omega t)H_1(\omega t)\bigg),
\end{equation}
where $J_n(x)$ denotes the Bessel function of order $n$ of the first kind and $H_n$ denotes the Struve function of order $n$ (see \cite[Chapter 11]{AS}).
\end{thm}
\begin{pf}
By employing the result in \cite[Lemma 4.1]{G}, the following relation holds: for sufficiently large $L$,
\begin{equation}\label{BesselInfinity}
  \sum_{l=1}^{L}e^{i\omega\mathbf{d}_l\cdot(\mathbf{x-z})}\approx \int_{\mathbb{S}^1}e^{i\omega\mathbf{d}\cdot(\mathbf{x-z})}dS(\mathbf{d})=2\pi J_0(\omega|\mathbf{x-z}|).
\end{equation}
Let $K\longrightarrow\infty$; then, applying an indefinite integral of the Bessel function (see \cite[page 3]{R}),
\[\int J_0(t)dt=tJ_0(t)+\frac{t\pi}{2}\bigg(J_1(t)H_0(t)-J_0(t)H_1(t)\bigg),\]
yields
\begin{align*}
  \sum_{k=1}^{K}d_T\mathbb{E}_\eps(\mathbf{z};\omega_k)&\approx2\pi\sum_{k=1}^{K}\int_\sigma (\eps-\eps_0)J_0(\omega_k|\mathbf{x-z}|)d\sigma(\mathbf{x})\approx2 \pi\int_\sigma\int_{\omega_1}^{\omega_K}(\eps-\eps_0)J_0(\omega|\mathbf{x-z}|)d\omega d\sigma(\mathbf{x})\\
  &=2\pi\int_\sigma(\eps-\eps_0)\bigg(\Lambda(t;\omega_K)-\Lambda(t;\omega_1)\bigg)d\sigma(\mathbf{x}),
\end{align*}
where function $\Lambda(t;\omega)$ is given by (\ref{FunctionLambda}). Hence, we can obtain the structure of $\mathbb{E}_3(\mathbf{z};K)$ via the above identity. Similarly, the structure of $\mathbb{E}_4(\mathbf{z};K)$ can be identified.
\end{pf}

\begin{thm}\label{TheoremMF3}Assume that the number of incident directions $L$ and frequency $\omega_K$ are sufficiently large enough, and $K$ is infinite ($K\longrightarrow\infty$); then, (\ref{MultiFrequencyTopologicalDerivative}) becomes
\[\mathbb{E}(\mathbf{z};K)\approx\frac{1}{2}\bigg(\frac{\mathbb{E}_5(\mathbf{z};K)}{\max|\mathbb{E}_5(\mathbf{z};K)|} +\frac{\mathbb{E}_6(\mathbf{z};K)}{\max|\mathbb{E}_6(\mathbf{z};K)|}\bigg),\]
where
\begin{align*}
  \mathbb{E}_5(\mathbf{z};K)&=\int_\sigma(\eps-\eps_0)\frac{2\pi}{|\mathbf{x-z}|}d\sigma(\mathbf{x})\\
  \mathbb{E}_6(\mathbf{z};K)&=\int_\sigma\bigg[2\bigg(\frac{1}{\mu}-\frac{1}{\mu_0}\bigg)\mathbf{d}_l\cdot\mathbf{t}(\mathbf{x}) +2\bigg(\frac{1}{\mu_0}-\frac{\mu}{\mu_0^2}\bigg)\mathbf{d}_l\cdot\mathbf{n}(\mathbf{x})\bigg]\frac{2\pi}{|\mathbf{x-z}|}d\sigma(\mathbf{x}).
\end{align*}
\end{thm}
\begin{pf}
By applying (\ref{BesselInfinity}), we can say that if $\omega_K\longrightarrow\infty$ and $K\longrightarrow\infty$, then,
\[\sum_{k=1}^{K}d_T\mathbb{E}_\eps(\mathbf{z};\omega_k)\approx2\pi\sum_{k=1}^{K}\int_\sigma(\eps-\eps_0) J_0(\omega_k|\mathbf{x-z}|)d\sigma(\mathbf{x})\approx2\pi\int_\sigma\int_0^{\infty}(\eps-\eps_0)J_0(\omega|\mathbf{x-z}|)d\omega d\sigma(\mathbf{x}).\]
Since following infinite integral of the Bessel function formula holds (see \cite[formula 11.4.17 (page 486)]{AS}),
\begin{equation}\label{InfiniteIntegralBessel}
  \int_0^\infty J_n(t)dt=1
\end{equation}
for $\mathfrak{Re}(n)>-1$. Then, applying change of variable $\omega|\mathbf{x-z}|=t$ in (\ref{InfiniteIntegralBessel}) for $n=0$ yields
\[\int_0^{\infty}J_0(\omega|\mathbf{x-z}|)d\omega=\int_0^{\infty}\frac{J_0(t)}{|\mathbf{x-z}|}dt=\frac{1}{|\mathbf{x-z}|}.\]
Thus, $d_T\mathbb{E}_\eps(\mathbf{z};\omega_k)$ becomes
\[d_T\mathbb{E}_\eps(\mathbf{z};\omega_k)\approx \int_\sigma(\eps-\eps_0)\frac{2\pi}{|\mathbf{x-z}|}d\sigma(\mathbf{x}),\]
and by taking the maximum value, we can obtain the desired result.
Similarly, $d_T\mathbb{E}_\mu(\mathbf{z};\omega_k)$ can be written as
\[d_T\mathbb{E}_\mu(\mathbf{z};\omega_k)\approx \int_\sigma\bigg[2\bigg(\frac{1}{\mu}-\frac{1}{\mu_0}\bigg)\mathbf{d}_l\cdot\mathbf{t}(\mathbf{x}) +2\bigg(\frac{1}{\mu_0}-\frac{\mu}{\mu_0^2}\bigg)\mathbf{d}_l\cdot\mathbf{n}(\mathbf{x})\bigg]\frac{2\pi}{|\mathbf{x-z}|}d\sigma(\mathbf{x}).\]
\end{pf}

\begin{figure}[!ht]
\begin{center}
\includegraphics[width=0.99\textwidth]{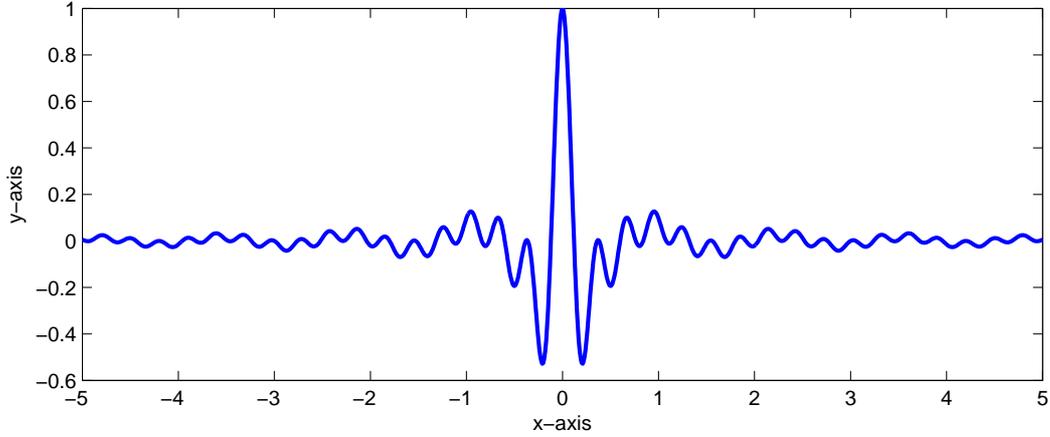}
\caption{Graph of $y=j_0(ax)\cos(bx)$ for $a=2$ and $b=10$.}\label{sincos}
\end{center}
\end{figure}

\begin{figure}[!ht]
\begin{center}
\includegraphics[width=0.99\textwidth]{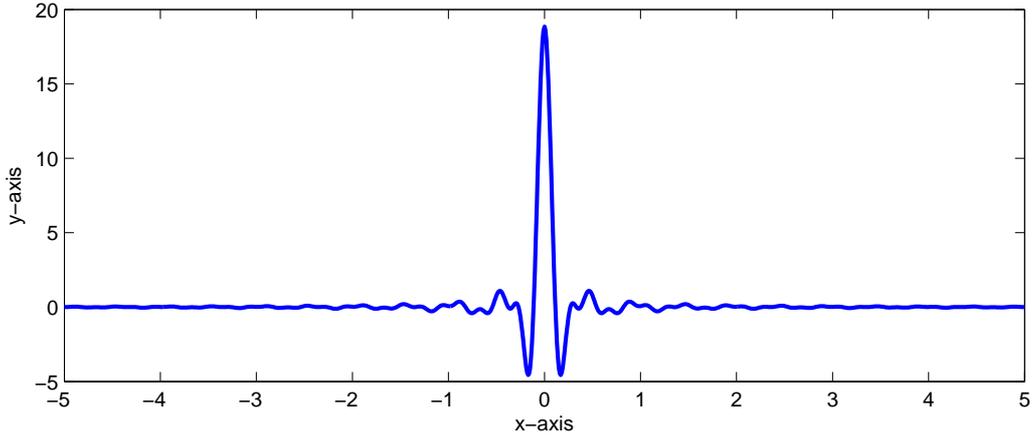}
\caption{Graph of $y=\Lambda(x;\omega_K)-\Lambda(x;\omega_1)$ for $\omega_K=\frac{2\pi}{0.2}$ and $\omega_1=\frac{2\pi}{0.5}$.}\label{Lambda}
\end{center}
\end{figure}

On the basis of Theorems \ref{TheoremMF1}, \ref{TheoremMF2}, and \ref{TheoremMF3}, we can explore some properties of normalized multi-frequency topological derivative imaging function (\ref{MultiFrequencyTopologicalDerivative}), summarized as follows:
\begin{enumerate}
  \item $j_0(\mathbf{x})$ and $\cos\mathbf{x}$ reach their maximum value $1$ at $\mathbf{x}=0$ and $\mathbf{x}=2n\pi$, respectively, for $n=0,\pm1,\pm2,\cdots$. Therefore, $\mathbb{E}(\mathbf{z};K)$ plots its maximum value at $\mathbf{z}$, which satisfies
      \[\xi_1\mathbf{d}_l\cdot(\mathbf{x-z})=0\quad\mbox{and}\quad\xi_2\mathbf{d}_l\cdot(\mathbf{x-z})=2n\pi\]
      for $n=0,\pm1,\pm2,\cdots$. This implies that points of magnitude $1$ (or close to $1$) will appear at $\mathbf{z}=\mathbf{x}$, i.e., along the unknown supporting curve $\sigma$. Moreover, since
      \[\lim_{x\to\infty}\frac{\sin ax}{ax}\cos bx\longrightarrow0,\]
      $\mathbb{E}(\mathbf{z};K)$ plots $0$ when $\mathbf{z}$ is far away from $\mathbf{x}$.
  \item $\Lambda(x;\omega_K)-\Lambda(x;\omega_1)$ has properties similar to those of $j_0(x)\cos x$, except for less oscillation. Hence, $\mathbb{E}(\mathbf{z};K)$ plots its maximum value at $\mathbf{z}=\mathbf{x}\in\sigma$.
  \item $j_0(ax)\cos(bx)$ and $\Lambda(x;\omega_K)-\Lambda(\mathbf{x};\omega_1)$ have their minimum values at two points $x_1$ and $x_2$, symmetric with respect to $x$, refer to Figures \ref{sincos} and \ref{Lambda}, respectively. This implies that the map of $\mathbb{E}(\mathbf{z};K)$ contains its minimum values in the neighborhood of $\sigma$ so that the location of the supporting curve is clearly identified by looking at points of maximum and minimum values.
  \item Applying multi-frequency (i.e., $K$ is sufficiently large enough) will guarantee a better imaging result than single frequency (i.e., $K=1$). Moreover, it is expected that applying a postprocessing operator introduced in \cite{AGJK} will yields a better result.
  \item The map of $\mathbb{E}(\mathbf{z};K)$ accurately yields the location of $\mathbf{z}=\mathbf{x}\in\sigma$ when we apply a large number of $K$ and $L$.
\end{enumerate}

\section{Numerical results and discussions}\label{Sec4}
\subsection{General configuration of numerical simulations}
Some numerical simulation results are presented herein. For simplicity, we consider the dielectric permittivity contrast case only. The homogeneous domain $\Omega$ is chosen as a unit circle centered at the origin in $\mathbb{R}^2$, and three $\sigma_j$ specify the thin inclusions $\Gamma_j$ as
\begin{flalign*}
  \sigma_1&=\set{(s-0.2,-0.5s^2+0.5)~:~s\in[-0.5,0.5]}&(\mbox{curve with constant curvature})\\
  \sigma_2&=\set{(s+0.2,s^3+s^2-0.6)~:~s\in[-0.5,0.5]}&(\mbox{curve with nonconstant curvature})\\
  \sigma_3&=\set{(s,0.5s^2+0.1\sin(3\pi(s+0.7))):s\in[-0.7,0.7]}.&(\mbox{oscillating curve})
\end{flalign*}
The thickness $h$ of the thin inclusion $\Gamma_j$ is set to $0.02$, and parameters $\eps_0$, $\mu_0$ are chosen as $1$. Let $\eps_j$ and $\mu_j$ for $j=1,2,3$ denote the permittivity and permeability of $\Gamma_j$, respectively. The applied frequency is selected as $\omega_k=\frac{2\pi}{\lambda_k}$ at wavelength $\lambda_k$, $k=1,2,\cdots,K$ and $L=4$ different incident directions
\[\mathbf{d}_l:=\bigg(\cos\frac{2(l-1)\pi}{L},\sin\frac{2(l-1)\pi}{L}\bigg),\quad l=1,2,\cdots,L,\]
have chosen. In order to show the robustness of the proposed algorithm, a white Gaussian noise with $15$dB signal-to-noise ratio (SNR) added to the unperturbed boundary data $u^{(l)}(\mathbf{x};\omega_k)$ via a standard MATLAB command `awgn'. Throughout this section, only both permittivity and permeability contrast case is considered, and we select $\eps_j=\mu_j=5$ for $j=1,2,$ and $3$.

\subsection{Numerical results and discussions}
\begin{figure}
\begin{center}
\includegraphics[width=0.49\textwidth]{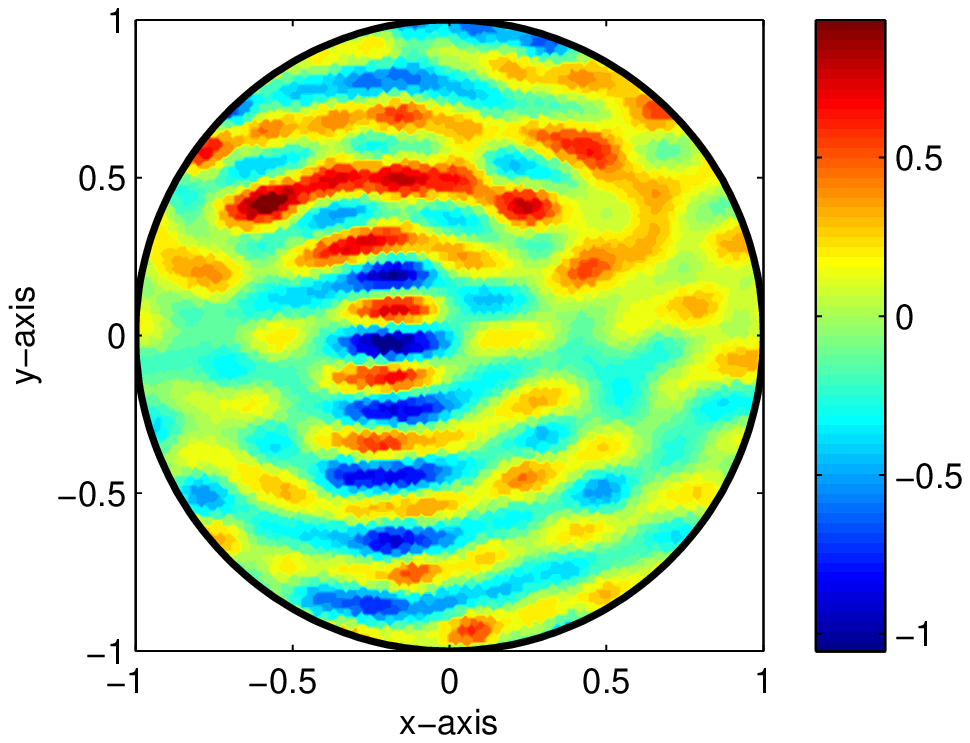}
\includegraphics[width=0.49\textwidth]{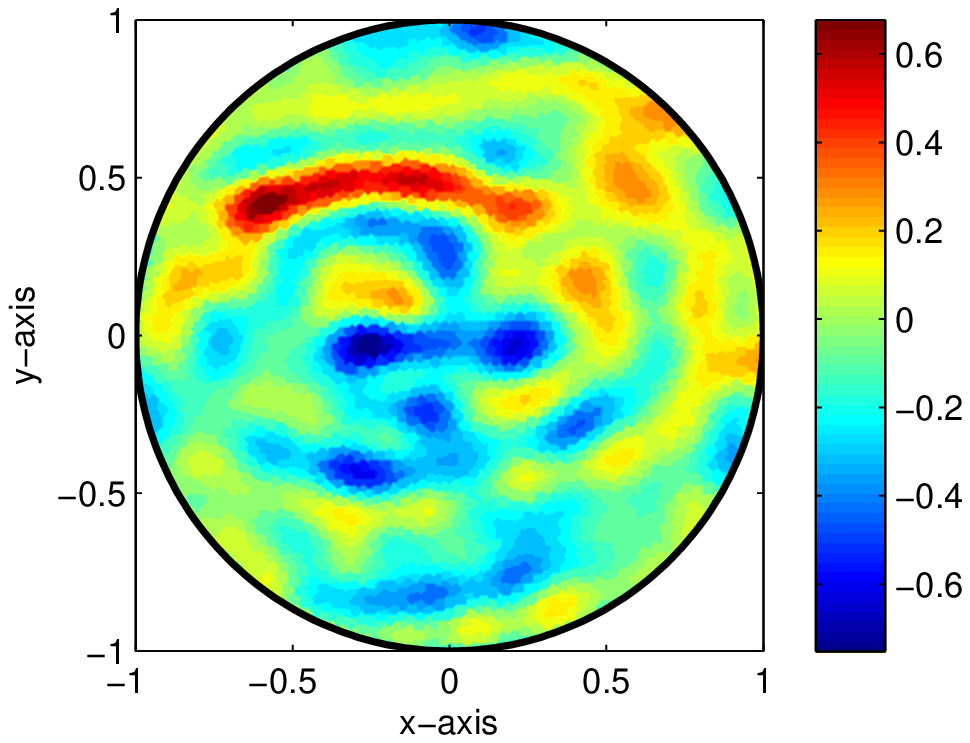}\\
\includegraphics[width=0.49\textwidth]{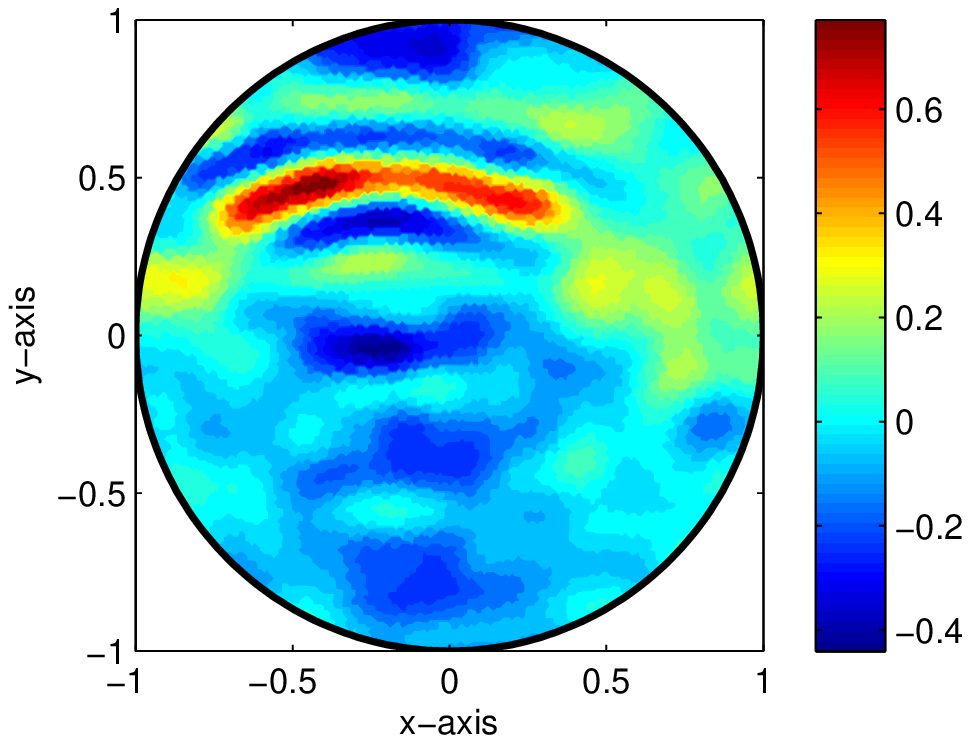}
\includegraphics[width=0.49\textwidth]{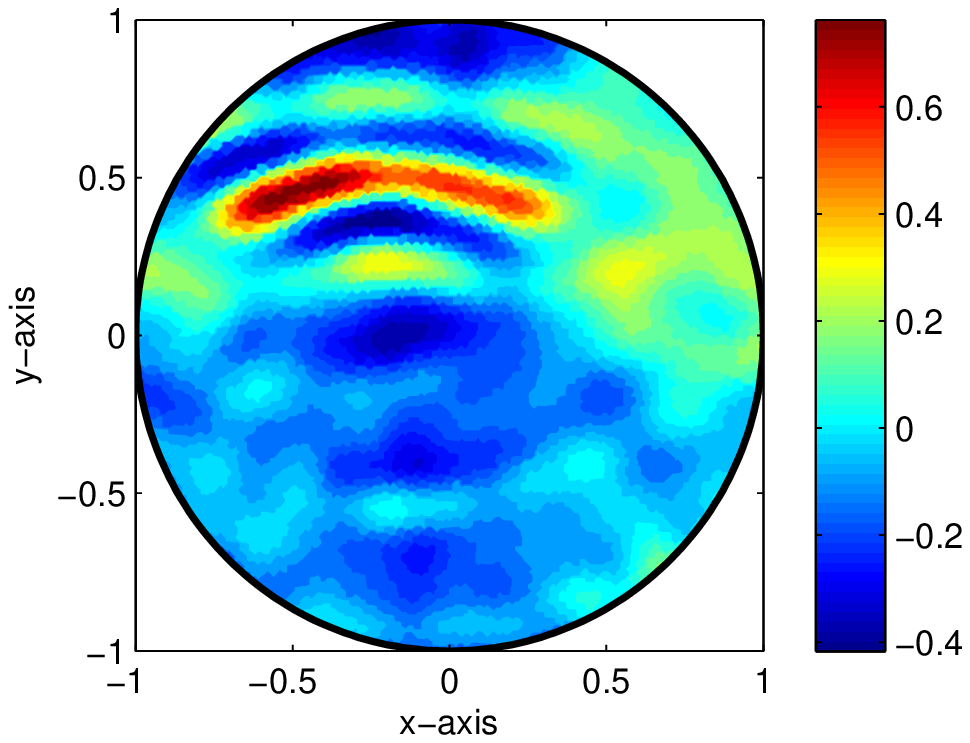}
\caption{Maps of $\mathbb{E}(\mathbf{z};K)$ for $L=4$, $K=1$ (top-left), $K=5$ (top-right), $K=10$ (bottom-left) and $K=16$ (bottom right) when the thin inclusion is $\Gamma_1$.}\label{Gamma1}
\end{center}
\end{figure}

First, let us consider the influence of the number of frequencies $K$. For this purpose, we choose a thin inclusion $\Gamma_1$ and compare maps of $\mathbb{E}(\mathbf{z};K)$ for $K=1,5,10,$ and $16$. From the results in Figure \ref{Gamma1}, it is difficult to recognize the shape of $\Gamma_1$ when we apply $K=1$ or $K=5$ because so many unexpected points of large magnitude are distributed on $\Omega\backslash\Gamma_1$. However, when we apply sufficiently large $K$, it is easy to recognize the shape of $\Gamma_1$. Based on the obtained image, $K=16$ is a good choice; hence, we will adopt $K=16$ different frequencies in this section. It is interesting to observe that when $K$ increases, the points of minimum value of $\mathbb{E}(\mathbf{z};K)$ appear in the neighborhood of $\Gamma_1$.

\begin{figure}
\begin{center}
\includegraphics[width=0.49\textwidth]{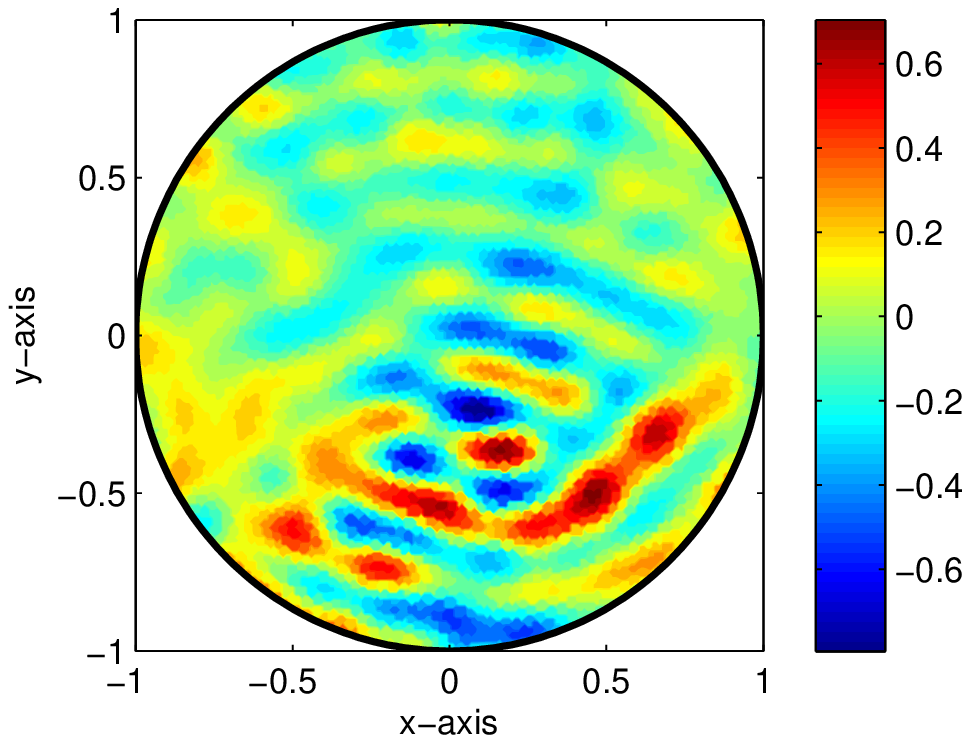}
\includegraphics[width=0.49\textwidth]{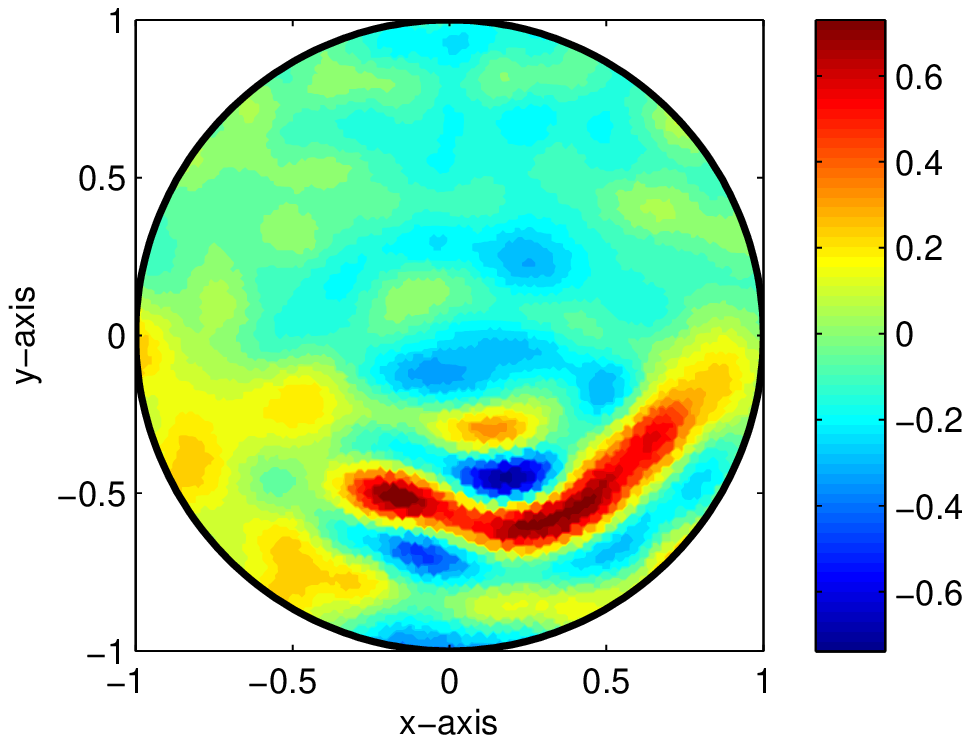}
\caption{Maps of $\mathbb{E}(\mathbf{z};K)$ for $L=4$ with $K=4$ (left) and $K=16$ (right) when the thin inclusion is $\Gamma_2$.}\label{Gamma2}
\end{center}
\end{figure}

Maps of $\mathbb{E}(\mathbf{z};K)$ are shown in Figure \ref{Gamma2} when the thin inclusion is $\Gamma_2$. Similar to the imaging of $\Gamma_1$, we can identify $\Gamma_2$ when the value $K$ is sufficiently large.

Let us apply the imaging function to $\Gamma_3$ under the same configuration as the above examples. Although only four points of $\Gamma_3$ are clearly identified, $\mathbb{E}(\mathbf{z};K)$ offers an acceptable result for an oscillating inclusion by comparing the result in \cite[Figure 5]{P3}.

\begin{figure}
\begin{center}
\includegraphics[width=0.49\textwidth]{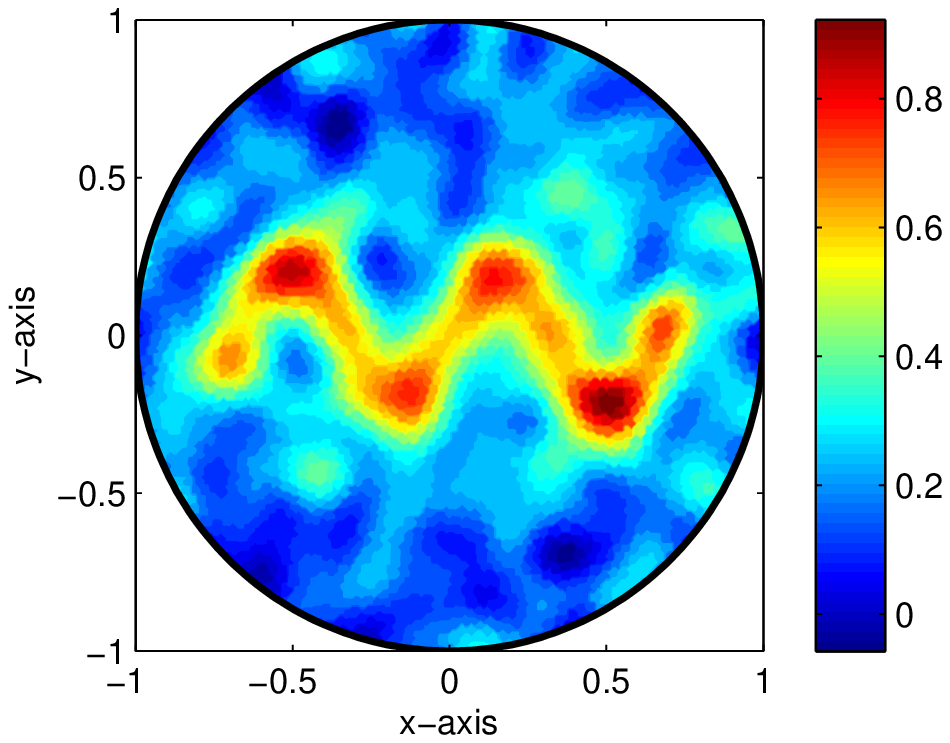}
\includegraphics[width=0.49\textwidth]{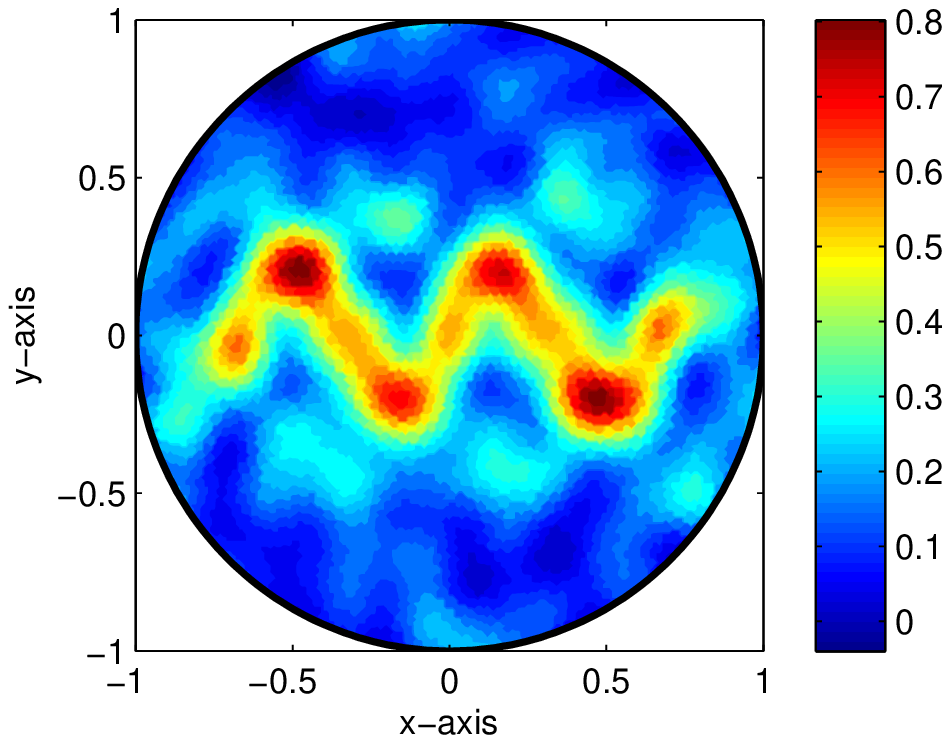}
\caption{Maps of $\mathbb{E}(\mathbf{z};K)$ for $L=4$ with $K=4$ (left) and $K=16$ (right) when the thin inclusion is $\Gamma_3$.}\label{Gamma3}
\end{center}
\end{figure}

\begin{figure}
\begin{center}
\includegraphics[width=0.49\textwidth]{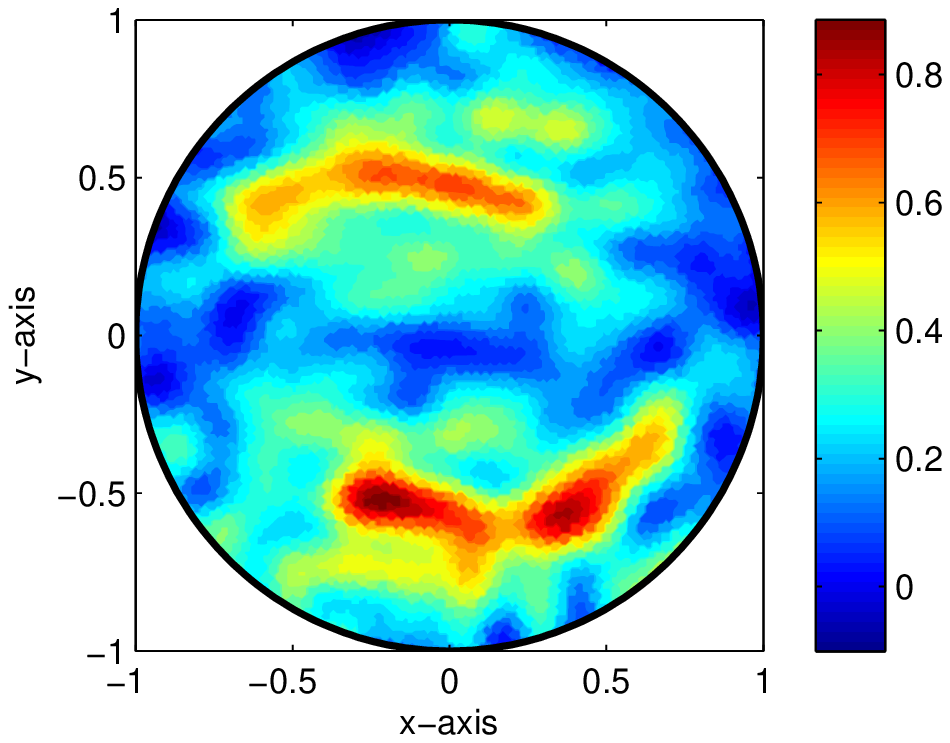}
\includegraphics[width=0.49\textwidth]{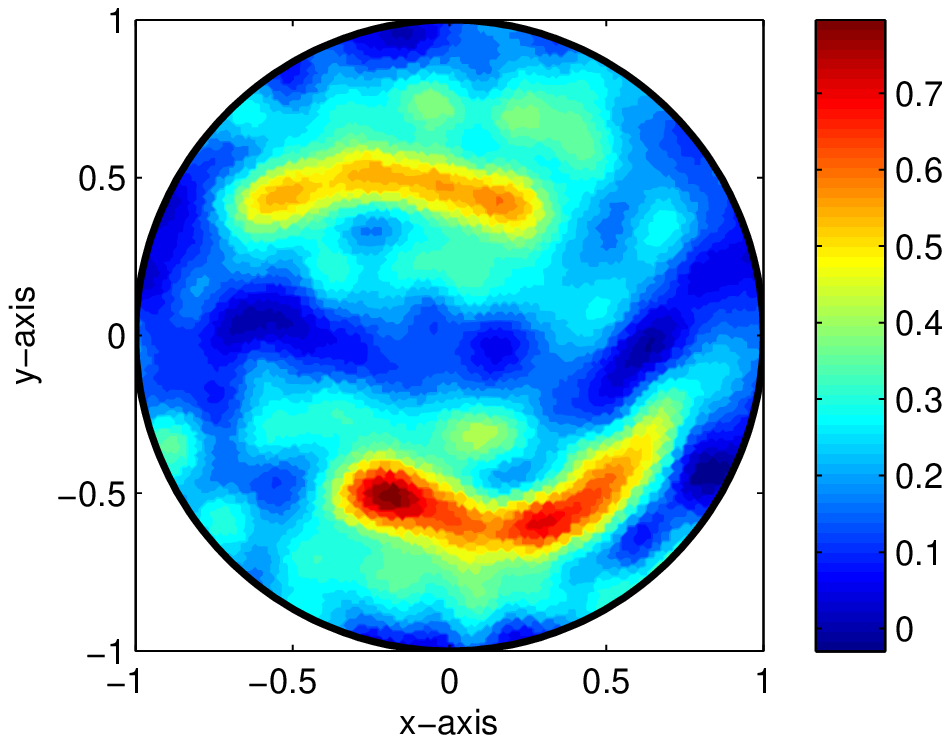}
\caption{Maps of $\mathbb{E}(\mathbf{z};K)$ for $L=4$ with $K=4$ (left) and $K=16$ (right) when the thin inclusion is $\Gamma_{\mbox{\tiny M}}$.}\label{GammaM1}
\end{center}
\end{figure}

One advantage of topological derivative is its straightforward application to the imaging of multiple inclusions. Figure \ref{GammaM1} shows the map of $\mathbb{E}(\mathbf{z};K)$ for imaging multiple thin inclusions $\Gamma_{\mbox{\tiny M}}=\Gamma_{\mbox{\tiny M1}}\cup\Gamma_{\mbox{\tiny M2}}=\Gamma_1\cup\Gamma_2$ with $\eps_1=\eps_2=5$ and $\mu_1=\mu_2=5$. Unlike to the previous single inclusion cases, although the existence of two inclusions can be recognized, it is difficule to identify their true shape.

\begin{figure}
\begin{center}
\includegraphics[width=0.49\textwidth]{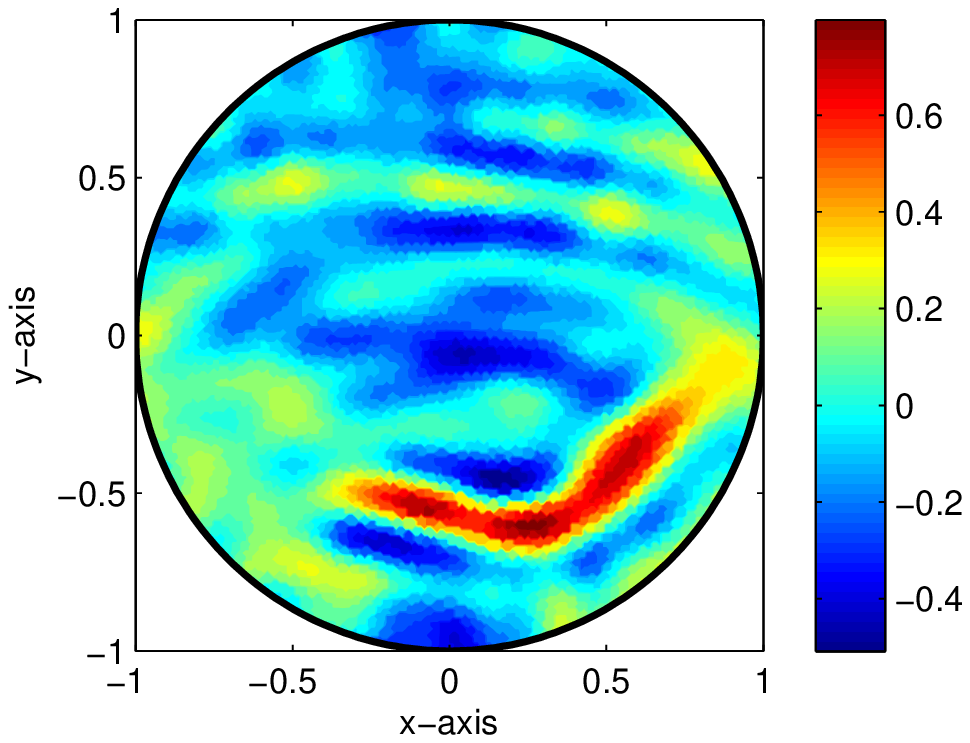}
\includegraphics[width=0.49\textwidth]{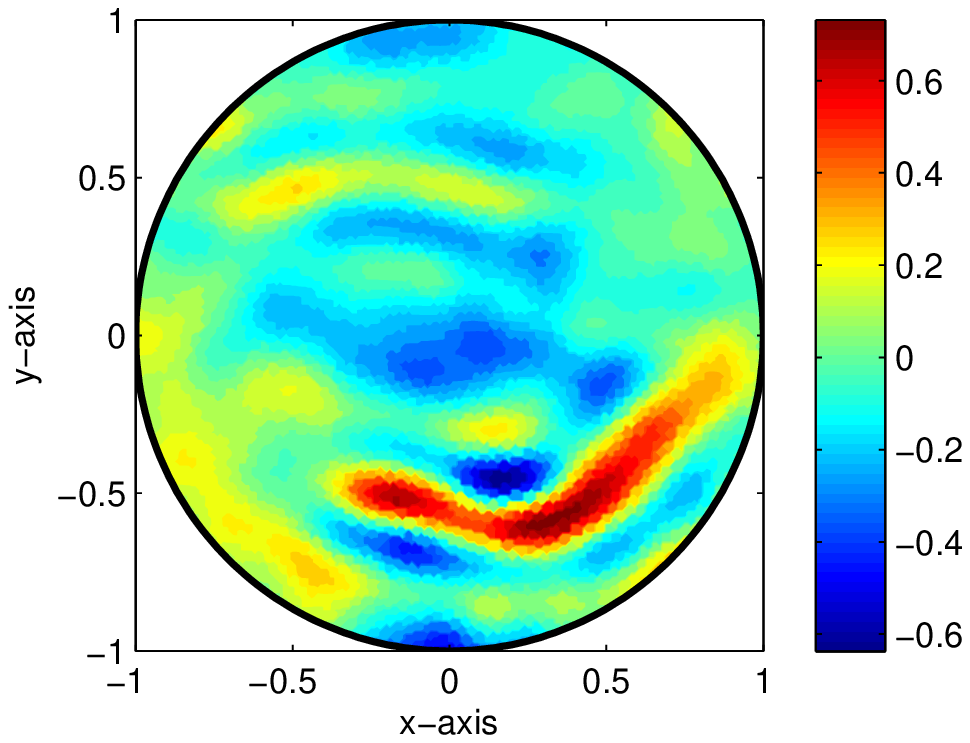}
\caption{Maps of $\mathbb{E}(\mathbf{z};K)$ for $L=4$ with $K=4$ (left) and $K=16$ (right) when the thin inclusion is $\Gamma_{\mbox{\tiny M}}$.}\label{GammaM2}
\end{center}
\end{figure}

Figure \ref{GammaM2} shows the map of $\mathbb{E}(\mathbf{z};K)$ under the same configuration as the previous example, except for different material properties, $\eps_1=\mu_1=5$ and $\eps_2=\mu_2=10$. Note that in the existence of $M-$different thin inclusions, Theorem \ref{TheoremMF1} becomes
\begin{align*}
\mathbb{E}_1(\mathbf{z};K)=\sum_{l=1}^{L}\sum_{m=1}^{M}\int_{\sigma_m}&(\eps_m-\eps_0)j_0\bigg(\frac{\omega_K-\omega_1}{2}\mathbf{d}_l\cdot(\mathbf{x-z})\bigg) \cos\bigg(\frac{\omega_K+\omega_1}{2}\mathbf{d}_l\cdot(\mathbf{x-z})\bigg)d\sigma(\mathbf{x})\\
\mathbb{E}_2(\mathbf{z};K)=\sum_{l=1}^{L}\sum_{m=1}^{M}\int_{\sigma_m}&\bigg[2\bigg(\frac{1}{\mu_m}-\frac{1}{\mu_0}\bigg)\mathbf{d}_l\cdot\mathbf{t}(\mathbf{x}) +2\bigg(\frac{1}{\mu_0}-\frac{\mu_m}{\mu_0^2}\bigg)\mathbf{d}_l\cdot\mathbf{n}(\mathbf{x})\bigg]\\
&\times j_0\bigg(\frac{\omega_K-\omega_1}{2}\mathbf{d}_l\cdot(\mathbf{x-z})\bigg) \cos\bigg(\frac{\omega_K+\omega_1}{2}\mathbf{d}_l\cdot(\mathbf{x-z})\bigg)d\sigma(\mathbf{x}).
\end{align*}
Theorems \ref{TheoremMF2} and \ref{TheoremMF3} can be written in a similar manner. Hence, it is true that if an inclusion (here, $\Gamma_1$) has a much smaller value of permittivity or permeability than another (here, $\Gamma_2$), this inclusion does not significantly affect the scattered field, and as a consequence, the value of $\mathbb{E}(\mathbf{z};K)$ for $\mathbf{z}\in\Gamma_1$ will be smaller than $\mathbb{E}(\mathbf{z};K)$ for $\mathbf{z}\in\Gamma_2$.

\begin{figure}
\begin{center}
\includegraphics[width=0.49\textwidth]{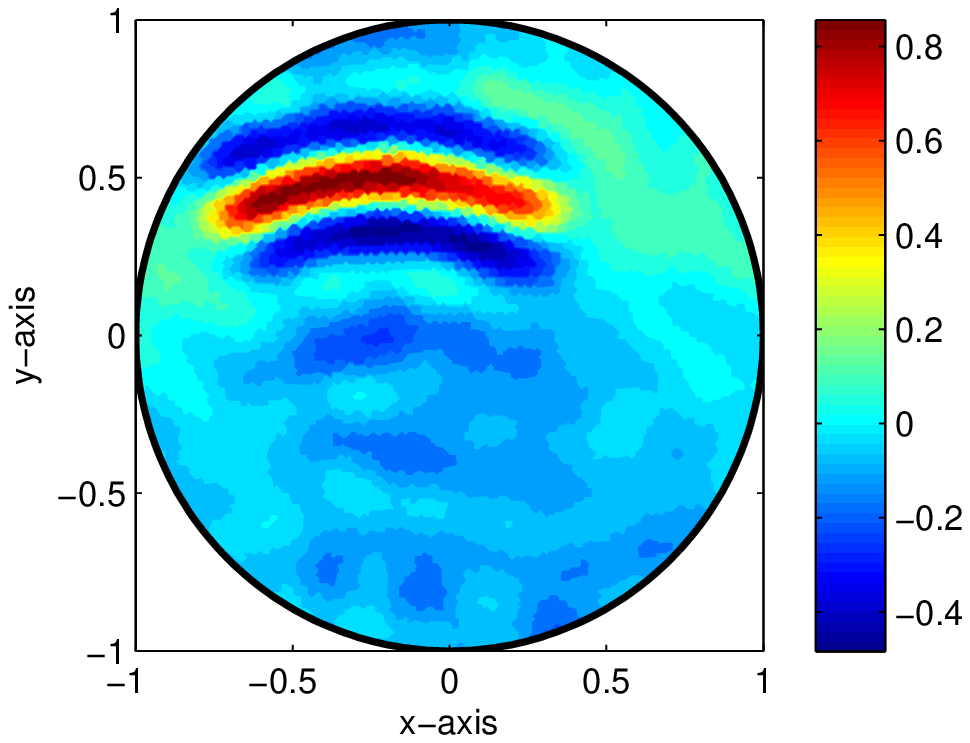}
\includegraphics[width=0.49\textwidth]{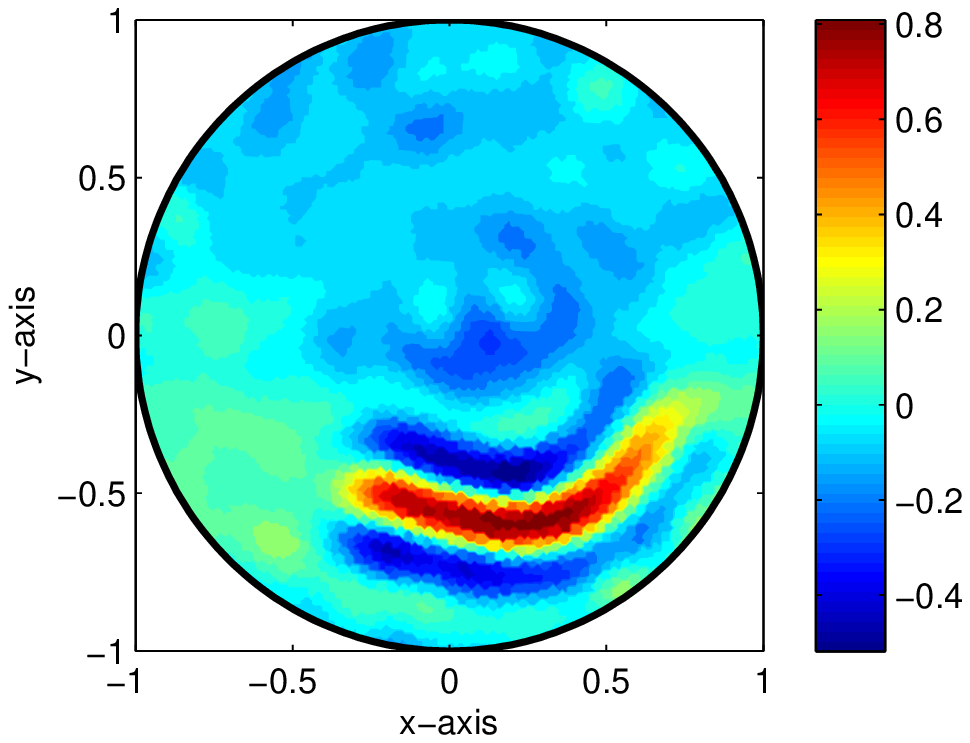}
\caption{Maps of $\mathbb{E}(\mathbf{z};K)$ for $L=16$, $K=16$ when the thin inclusion is $\Gamma_1$ (left) and $\Gamma_2$ (right).}\label{Gamma12-16}
\end{center}
\end{figure}

\begin{figure}
\begin{center}
\includegraphics[width=0.49\textwidth]{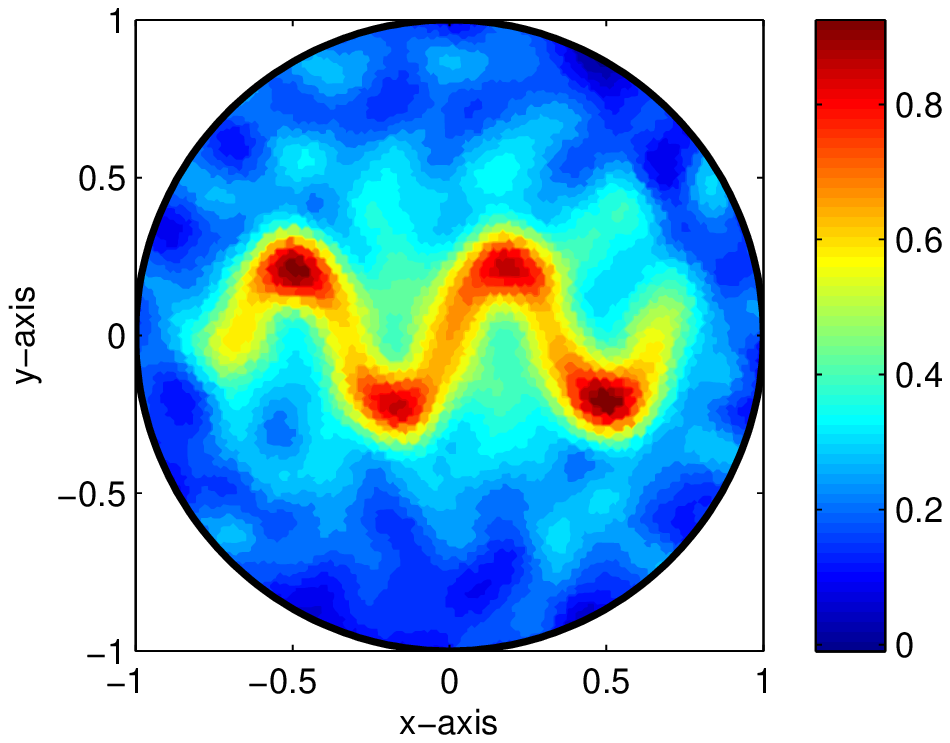}
\includegraphics[width=0.49\textwidth]{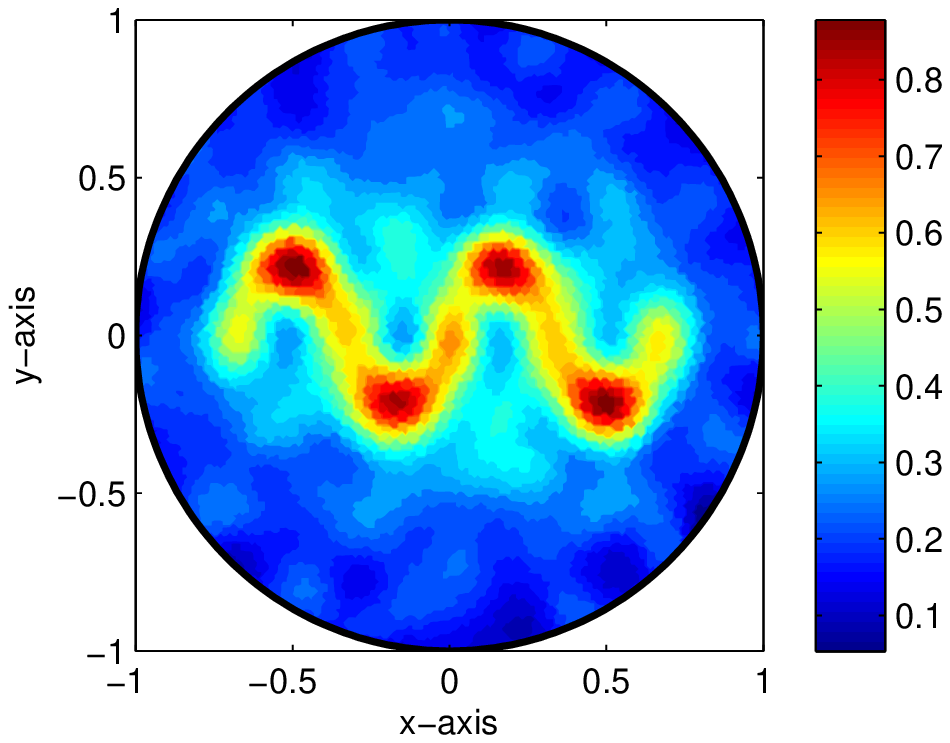}
\caption{Maps of $\mathbb{E}(\mathbf{z};K)$ for $L=16$ with $K=4$ (left) and $K=L=16$ (right) when the thin inclusion is $\Gamma_3$.}\label{Gamma3-16}
\end{center}
\end{figure}

An improvement can be realized by simply making $L$ as large as possible. Figures \ref{Gamma12-16} and \ref{Gamma3-16} are maps of $\mathbb{E}(\mathbf{z};K)$ for $L=16$ in the existence of a single thin inclusion. By comparing Figures \ref{Gamma1}, \ref{Gamma2} and \ref{Gamma3}, the shape of $\Gamma_j$ appears more accurate than the $L=4$ case. Note that if one can apply a large number of incident directions $L$, the number of applied frequencies $K$ can be reduced, refer to Figure \ref{Gamma3-16}. Figure \ref{GammaM1M2-16} shows the map of $\mathbb{E}(\mathbf{z};K)$ with $K=L=16$ in the existence of multiple thin inclusions. As expected, good imaging results are obtained.

\begin{figure}
\begin{center}
\includegraphics[width=0.49\textwidth]{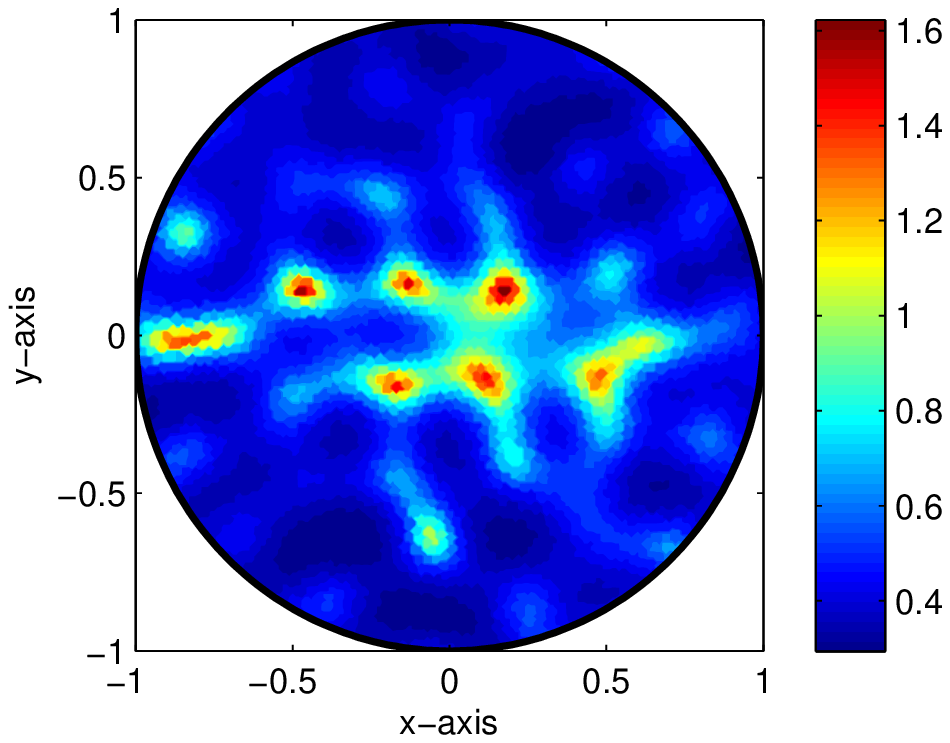}
\includegraphics[width=0.49\textwidth]{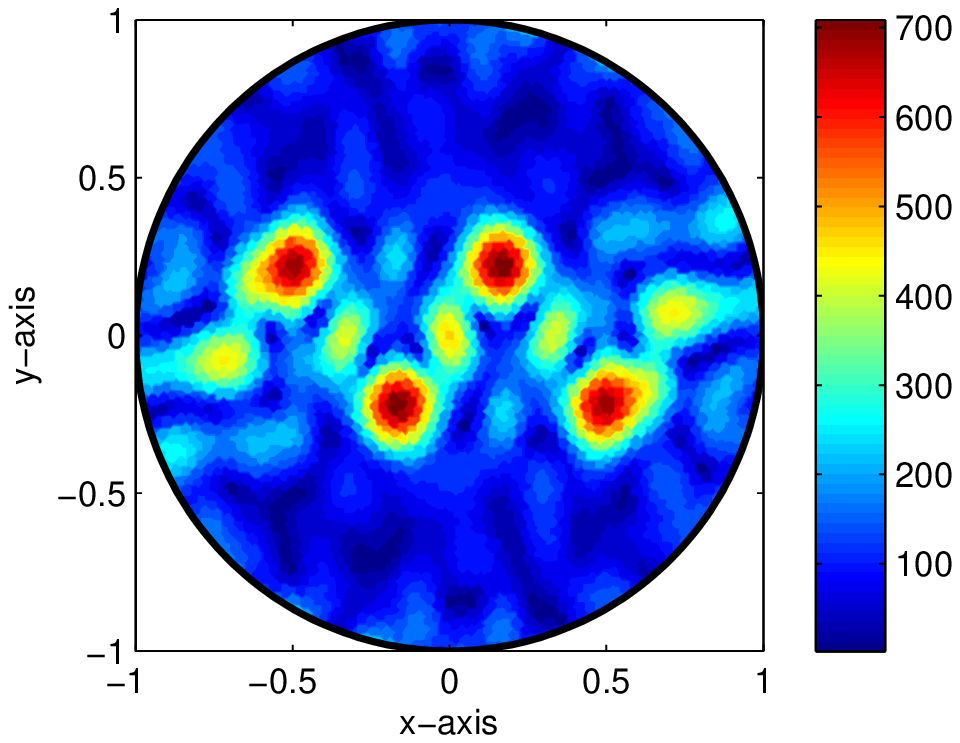}
\caption{Imaging result via MUSIC(left) at single frequency $\omega=\frac{2\pi}{0.5}$ and Kirchhoff migration (right) at multi-frequency when the thin inclusion is $\Gamma_3$.}\label{Gamma3-Compare}
\end{center}
\end{figure}

\begin{figure}
\begin{center}
\includegraphics[width=0.49\textwidth]{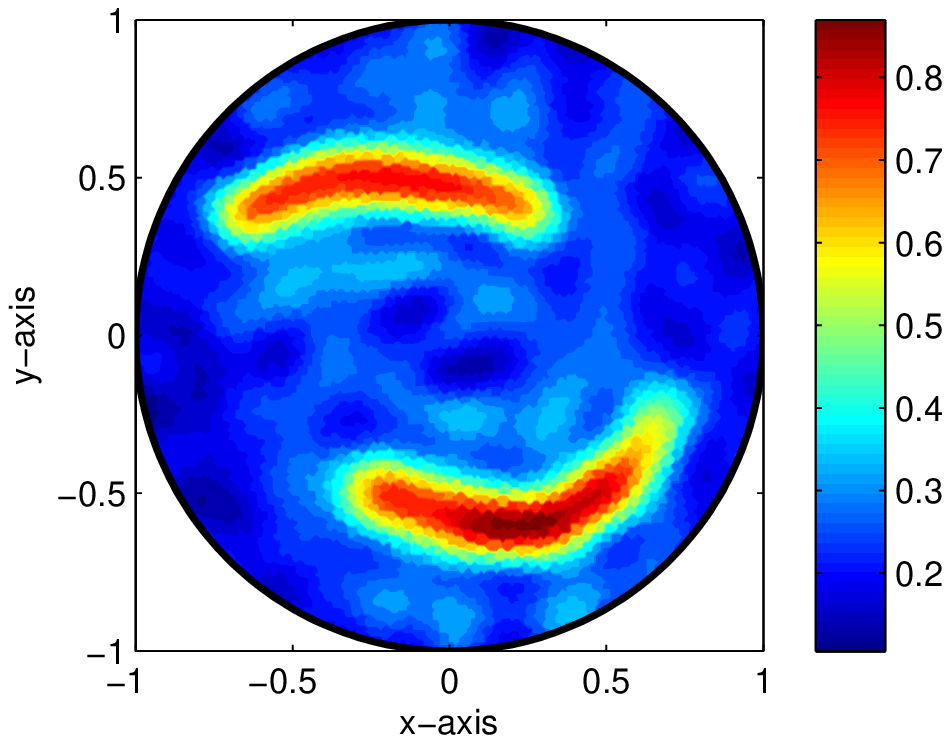}
\includegraphics[width=0.49\textwidth]{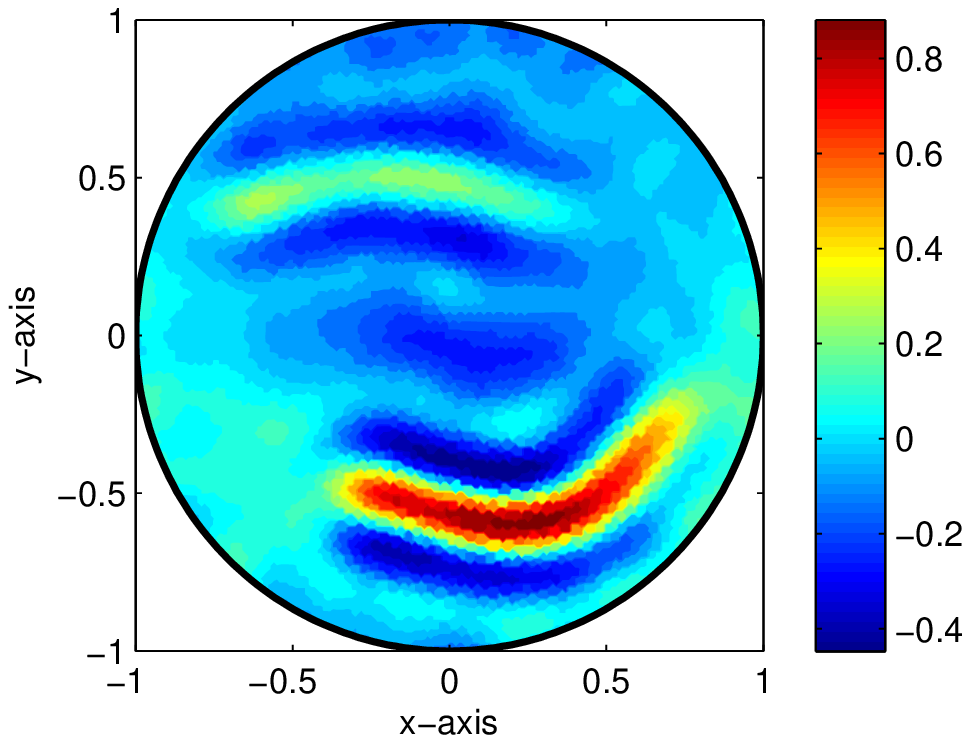}
\caption{Maps of $\mathbb{E}(\mathbf{z};K)$ for $K=L=16$ with same (left) and different (right) permittivities and permeabilities when the thin inclusion is $\Gamma_{\mbox{\tiny M}}$.}\label{GammaM1M2-16}
\end{center}
\end{figure}

Now, let us compare $\mathbb{E}(\mathbf{z};K)$ with two well-known non-iterative algorithms, MUltiple SIgnal Classification (MUSIC) and Kirchhoff migrations (see Appendix \ref{SecA} for corresponding algorithms). Figure \ref{Gamma3-Compare} shows the imaging result of MUSIC and Kirchhoff migrations for $L=16$ when the thin inclusion is $\Gamma_3$ without noisy data. From Figures \ref{Gamma3-16} and \ref{Gamma3-Compare}, we can observe that because of the small value of $L$\footnote{If the value of $L$ is sufficiently large enough, good result can be obtained, refer to \cite{P1,PL1,PL2}}, a good result cannot be obtained via MUSIC and Kirchhoff migrations, but $\mathbb{E}(\mathbf{z};K)$ yields a good result.

\subsection{Producing a good initial guess for applying iterative algorithms}
\begin{figure}
\begin{center}
\includegraphics[width=0.49\textwidth]{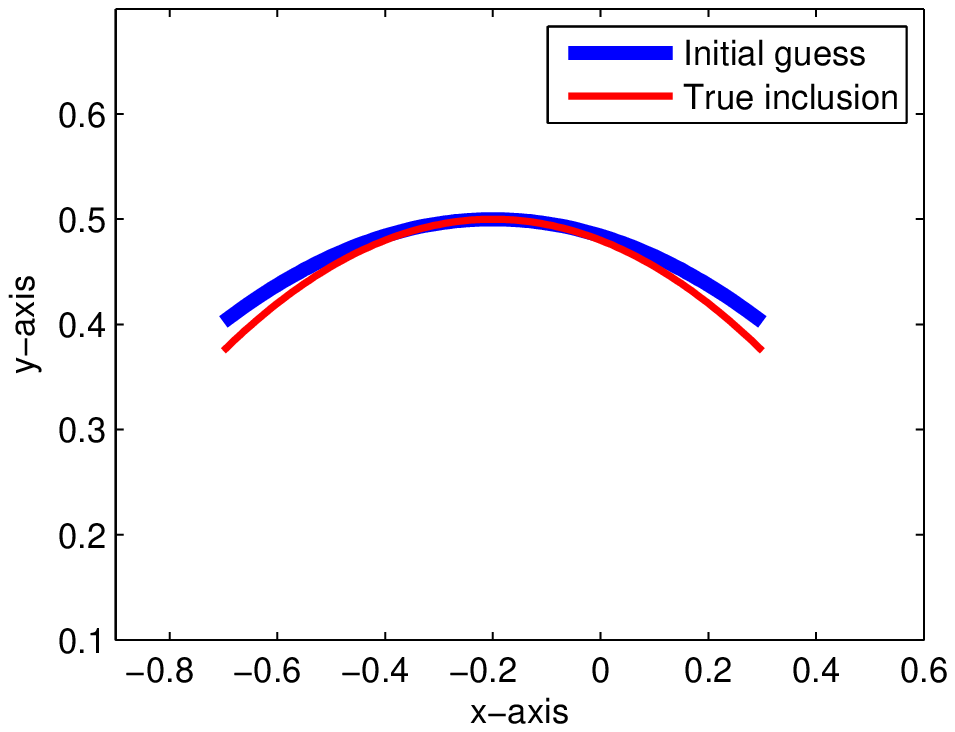}
\includegraphics[width=0.49\textwidth]{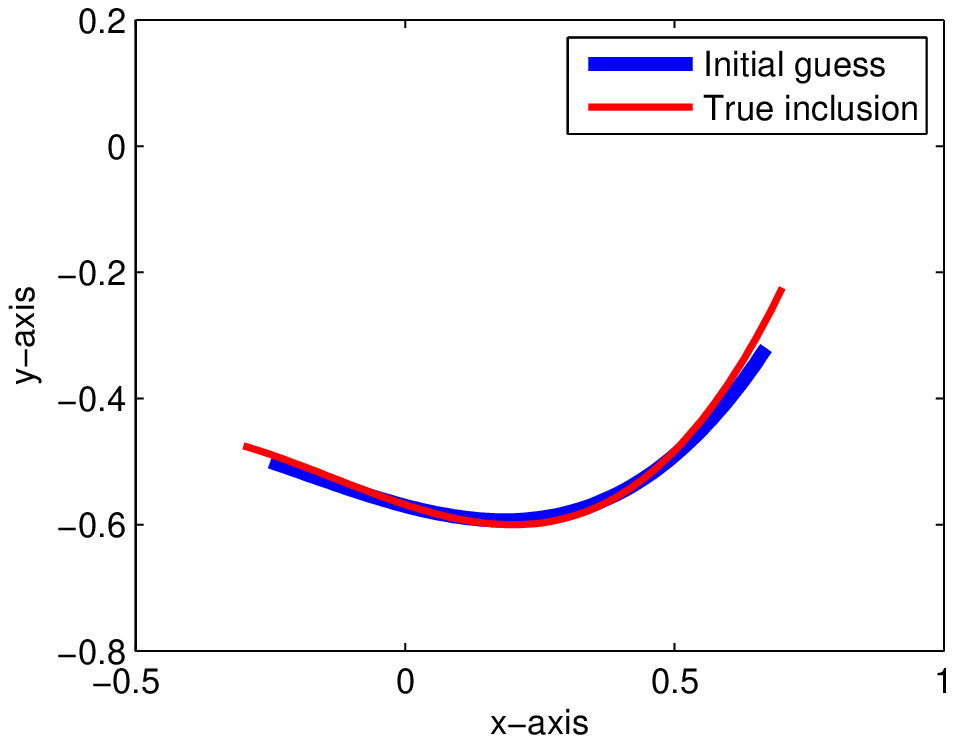}\\
\includegraphics[width=0.49\textwidth]{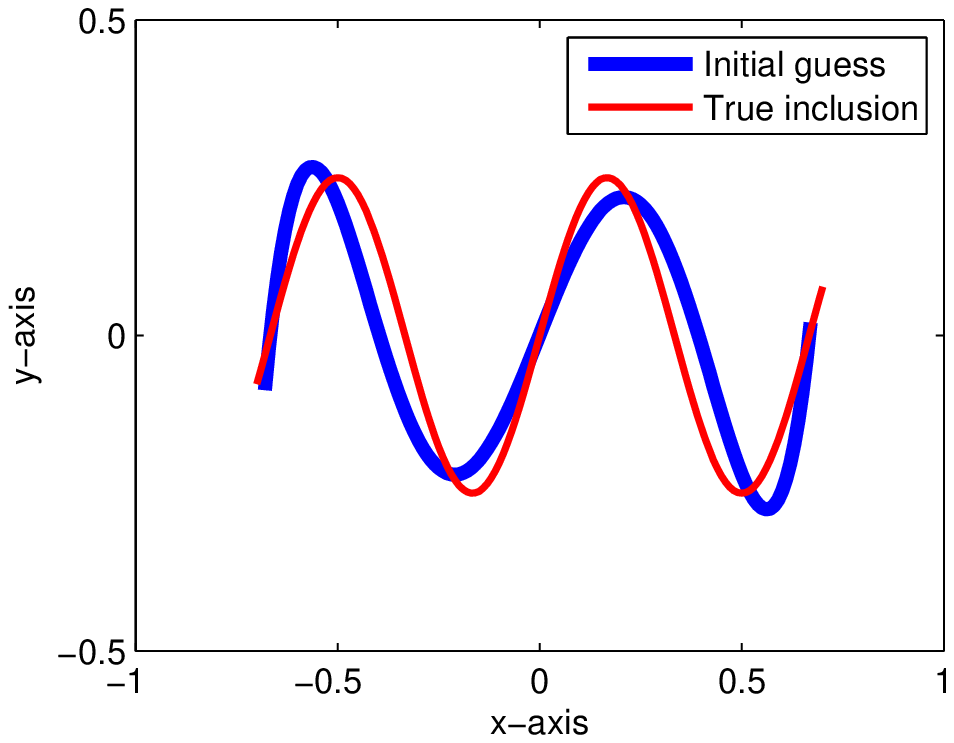}
\includegraphics[width=0.49\textwidth]{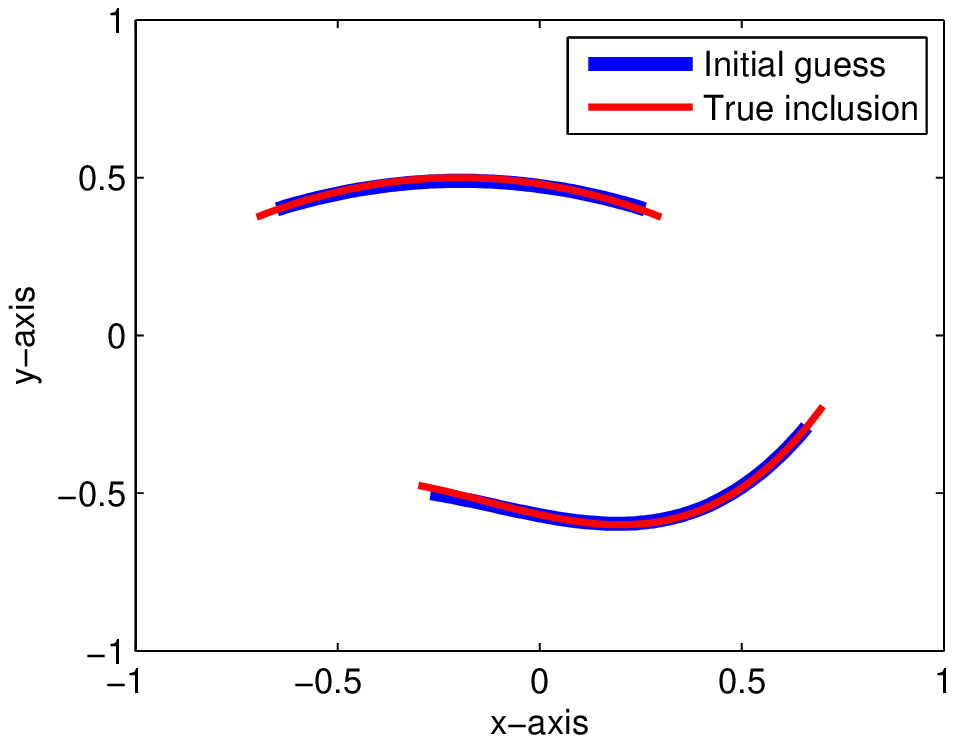}
\caption{Illustration of obtained initial guess $\sigma_1^{\mbox{\tiny init}}$ (top, left), $\sigma_2^{\mbox{\tiny init}}$ (top, right), $\sigma_3^{\mbox{\tiny init}}$ (bottom, left), and $\sigma_{\mbox{\tiny M1}}^{\mbox{\tiny init}}\cup\sigma_{\mbox{\tiny M2}}^{\mbox{\tiny init}}$ (bottom, right)}\label{GammaInitial}
\end{center}
\end{figure}

From the results presented in the previous section, we can generate a good initial guess for iterative based reconstruction algorithm \cite{ADIM,CR,DL,PL4}. Borrowing the basic idea of \cite{K}, we assume that the supporting curve $\sigma_j$ can be represented as follows
\[\sigma_j=\set{z_j(s):s\in[a_j,b_j]},\]
where $z_j:[a_j,b_j]\longrightarrow\mathbb{R}^2$ is of the form
\[z_j(s)=\left(s,\sum_{p=0}^{q}c_p T_p(s)\right),\quad s\in[a_j,b_j].\]
Here, $T_p(s)$ denotes the Chebyshev polynomials of the first kind, defined by the recurrence relation
\begin{align*}
  T_0(s)&=1\\
  T_1(s)&=s\\
  T_{p+1}(s)&=2sT_p(s)-T_{p-1}(s).
\end{align*}
From the numerical experience in \cite[Section 7]{K}, we use $q=5$ polynomials $T_p(s)$, $p=1,2,\cdots,q$, in order to represent $\sigma_j$. The computed coefficients $c_p$ listed in Table \ref{CoefficientError}, and the corresponding curves $\sigma_j^{\mbox{\tiny init}}$ are shown in Figure \ref{GammaInitial}.
\begin{table}[!ht]
\begin{center}
\begin{tabular}{c||c|c|c|c|c|c||c|c|c}
\hline Curve&$c_0$&$c_1$&$c_2$&$c_3$&$c_4$&$c_5$&$\mathbb{N}_1(\omega)$&$\mathbb{N}_2(\omega)$&$\mathbb{N}_\infty(\omega)$\\
\hline\hline
$\sigma_1^{\mbox{\tiny init}}$&$\phantom{-}0.2891$&$-0.1563$&$-0.1963$&$0.0000$&$0.0000$&$0.0000$&$0.1021$&$0.1569
$&$0.7388$\\
$\sigma_2^{\mbox{\tiny init}}$&$-0.3673$&$\phantom{-}0.3198$&$\phantom{-}0.2027$&$0.1813$&$0.0000$&$0.0000$&$0.1396$&$0.2304$&$1.0716$\\
$\sigma_3^{\mbox{\tiny init}}$&$\phantom{-}0.0169$&$\phantom{-}5.7194$&$\phantom{-}0.0297$&$3.8071$&$0.0142$&$1.4649$&$0.2391$&$0.4095$&$2.5765$\\
\hline
$\sigma_{\mbox{\tiny M1}}^{\mbox{\tiny init}}$&$\phantom{-}0.2561$&$-0.1696$&$-0.2174$&$0.0000$&$0.0000$&$0.0000$&\multirow{2}*{$0.3251$}&\multirow{2}*{$0.4823$}&\multirow{2}*{$2.6507$}\\
$\sigma_{\mbox{\tiny M2}}^{\mbox{\tiny init}}$&$-0.4018$&$\phantom{-}0.5235$&$\phantom{-}0.1682$&$0.2555$&$0.0000$&$0.0000$& & &\\
\hline
\end{tabular}
\caption{\label{CoefficientError}Computed coefficients $a_p$ of Chebyshev polynomials of the first kind $T_p(s)$, $p=0,1,\cdots,5,$ and values of discrete norms $\mathbb{N}_1(\omega)$, $\mathbb{N}_2(\omega)$, and $\mathbb{N}_\infty(\omega)$ for $\omega=\frac{2\pi}{0.5}$.}
\end{center}
\end{table}

Let $u_{\mbox{\tiny true}}^{(l)}(\mathbf{x};\omega)$ and $u_{\mbox{\tiny comp}}^{(l)}(\mathbf{x};\omega)$ be the solution of (\ref{ForwardProblem}) in the existence of a true inclusion $\Gamma_j$ and initial guess $\Gamma_j^{\mbox{\tiny init}}$ with supporting curve $\sigma_j^{\mbox{\tiny init}}$, respectively. Then, due to the difference in shape of $\Gamma_j$ and $\Gamma_j^{\mbox{\tiny init}}$, we can define some discrete norms and evaluate them in order to investigate the fact that the obtained thin inclusion $\Gamma_j^{\mbox{\tiny init}}$ is close to the true one $\Gamma_j$: for $\mathbf{x}_n\in\partial\Omega$, $n=1,2,\cdots,N$,
\begin{align*}
  \mathbb{N}_1(\omega)&:=\frac{1}{L}\sum_{l=1}^{L}\|u_{\mbox{\tiny true}}^{(l)}(\mathbf{x};\omega)-u_{\mbox{\tiny comp}}^{(l)}(\mathbf{x};\omega)\|_{\ell^1(\partial\Omega)}=\frac{1}{L}\sum_{l=1}^{L}\sum_{n=1}^{N}|u_{\mbox{\tiny true}}^{(l)}(\mathbf{x}_n;\omega)-u_{\mbox{\tiny comp}}^{(l)}(\mathbf{x}_n;\omega)|,\\
  \mathbb{N}_2(\omega)&:=\frac{1}{L}\sum_{l=1}^{L}\|u_{\mbox{\tiny true}}^{(l)}(\mathbf{x};\omega)-u_{\mbox{\tiny comp}}^{(l)}(\mathbf{x};\omega)\|_{\ell^2(\partial\Omega)}=\frac{1}{L}\sum_{l=1}^{L}\left(\sum_{n=1}^{N}|u_{\mbox{\tiny true}}^{(l)}(\mathbf{x}_n;\omega)-u_{\mbox{\tiny comp}}^{(l)}(\mathbf{x}_n;\omega)|^2\right)^{\frac12},\\
  \mathbb{N}_\infty(\omega)&:=\frac{1}{L}\sum_{l=1}^{L}\|u_{\mbox{\tiny true}}^{(l)}(\mathbf{x};\omega)-u_{\mbox{\tiny comp}}^{(l)}(\mathbf{x};\omega)\|_{\ell^\infty(\partial\Omega)}=\frac{1}{L}\sum_{l=1}^{L}\max_{\mathbf{x}_n\in\partial\Omega} |u_{\mbox{\tiny true}}^{(l)}(\mathbf{x}_n;\omega)-u_{\mbox{\tiny comp}}^{(l)}(\mathbf{x}_n;\omega)|.
\end{align*}
Notice that in this paper, since $\Omega$ is a unit circle, $N=128$ different points $\mathbf{x}_n$ on the boundary $\partial\Omega$ are chosen as
\[\mathbf{x}_n=\left(\cos\frac{2n\pi}{N},\sin\frac{2n\pi}{N}\right)\quad\mbox{for}\quad n=1,2,\cdots,N.\]
In Table \ref{CoefficientError}, values of $\mathbb{N}_1(\omega)$, $\mathbb{N}_2(\omega)$, and $\mathbb{N}_\infty(\omega)$ for $\omega=\frac{2\pi}{0.5}$ are listed for thin inclusions $\Gamma_1$, $\Gamma_2$, $\Gamma_3$, and $\Gamma_{\mbox{\tiny M}}$. Obtained supporting curves $\sigma_j^{\mbox{\tiny init}}$ and corresponding values of discrete norms indicate that a good initial guess is obtained and it will be useful for performing complete shape reconstruction via an iterative algorithm, for example, level set method introduced in \cite{PL4}.

\section{Conclusion}\label{Sec5}
We investigated the applicability of a multi-frequency based topological derivative algorithm for the imaging of two-dimensional thin, penetrable inclusions embedded in a homogeneous domain. Various numerical results indicate that the proposed algorithm is stable even in the existence of random noise, and it is applicable for single and multiple inclusions. Although the shapes obtained via imaging results do not yield the complete shape of the inclusion with certainty, the iterative reconstruction algorithm can be successfully performed by employing them as an initial guess.

However, the proposed algorithm has some limitations; for example, it cannot be applied to the limited-view inverse problems in contrast to the Kirchhoff migration, refer to \cite{AGKPS,P1,P2,PL2}. Therefore, the supplement of a deficiency point of the proposed algorithm will be an interesting subject.

In this contribution, we considered the imaging of thin electromagnetic inclusions when measured boundary data is polluted by Gaussian random noise. We believe that the proposed algorithm can be applied for imaging when the measured data is distorted by random scatterers.

\appendix

\section{MUSIC algorithm and Kirchhoff migration}\label{SecA}
In this appendix, we briefly introduce the well-known MUSIC algorithm and Kirchhoff migration. More detailed discussion can be found in various literatures \cite{AGKPS,AILP,AK,AKLP,C,HSSZ,P1,P2,PL1,PL3} and references therein.

Same as in section \ref{Sec2}, let $u_{\mbox{\tiny tot}}^{(l)}(\mathbf{x};\omega)$ and $u_{\mbox{\tiny bac}}^{(l)}(\mathbf{x};\omega)$ denote the total and background solutions of (\ref{ForwardProblem}), respectively. Then, scattered field measured at boundary $\partial\Omega$ can be written as an asymptotic expansion formula (see \cite{BF} for instance),
\begin{align}
\begin{aligned}\label{AsymptoticExpansionFormula}
u_{\mbox{\tiny tot}}^{(l)}(\mathbf{y};\omega)-u_{\mbox{\tiny bac}}^{(l)}(\mathbf{y};\omega)=h\int_{\sigma}\bigg[&\nabla u_{\mbox{\tiny bac}}^{(l)}(\mathbf{x};\omega)\cdot\mathbb{M}(\sigma;\mathbf{x}) \cdot\nabla\mathcal{N}(\mathbf{x},\mathbf{y};\omega)\\
&+\omega^2(\eps-\eps_0)u_{\mbox{\tiny bac}}^{(l)}(\mathbf{x};\omega)\mathcal{N}(\mathbf{x},\mathbf{y};\omega)\bigg]d\sigma(\mathbf{x})+o(h),
\end{aligned}
\end{align}
where $\mathcal{N}(\mathbf{x},\mathbf{y};\omega)$ denotes the Neumann function for Helmholtz operator $\nabla^2+\omega^2\eps_0\mu_0$ in $\Omega$ corresponding to the Dirac delta function $-\delta(\mathbf{x},\mathbf{y})$ that satisfies
\begin{equation}\label{FunctionN}
\left\{\begin{array}{rcl}
\nabla^2\mathcal{N}(\mathbf{x},\mathbf{y};\omega) +\omega^2\eps_0\mu_0\mathcal{N}(\mathbf{x},\mathbf{y};\omega)=-\delta(\mathbf{x},\mathbf{y})&\mbox{in}&\Omega\\
\noalign{\medskip}\displaystyle\frac{\p \mathcal{N}(\mathbf{x},\mathbf{y};\omega)}{\p\boldsymbol{\nu}(\mathbf{x})}=0&\mbox{on}&\p\Omega,
\end{array}\right.
\end{equation}
and a symmetric matrix $\mathbb{M}(\sigma;\mathbf{x})$ is defined as follows: for $\mathbf{x}\in\sigma$, let $\mathbf{t}(\mathbf{x})$ and $\mathbf{n}(\mathbf{x})$ denote unit tangent and normal vectors to $\sigma$ at $\mathbf{x}$, respectively. Then
\begin{itemize}
  \item $\mathbb{M}(\sigma;\mathbf{x})$ has eigenvectors $\mathbf{t}(\mathbf{x})$ and $\mathbf{n}(\mathbf{x})$.
  \item The eigenvalue corresponding to $\mathbf{t}(\mathbf{x})$ is $2\left(\frac{1}{\mu}-\frac{1}{\mu_0}\right)$.
  \item The eigenvalue corresponding to $\mathbf{n}(\mathbf{x})$ is $2\left(\frac{1}{\mu_0}-\frac{\mu}{\mu_0^2}\right)$.
\end{itemize}

Now, let $\{\mathbf{d}_l\}_{l=1}^{L}\subset\mathbb{S}^1$ be a discrete finite set of incident directions and $\{\mathbf{d}_j^\bot\}_{j=1}^{L}\subset \mathbb{S}^1$ be the same number of observation directions. Then, applying asymptotic expansion formula (\ref{AsymptoticExpansionFormula}) and performing integration by parts, we can obtain the following normalized boundary measurements:
\begin{align*}
&\int_{\p\Omega}\bigg(u_{\mbox{\tiny tot}}^{(l)}(\mathbf{y};\omega)-u_{\mbox{\tiny bac}}^{(l)}(\mathbf{y};\omega)\bigg)\frac{\p u_{\mbox{\tiny bac}}^{(j)}(\mathbf{y};\omega)}{\p\boldsymbol{\nu}(\mathbf{y})}dS(\mathbf{y})\\
=&\int_{\p\Omega}h\int_{\sigma}\bigg[\nabla u_{\mbox{\tiny bac}}^{(l)}(\mathbf{x};\omega)\cdot\mathbb{M}(\sigma;\mathbf{x}) \cdot\nabla\mathcal{N}(\mathbf{x},\mathbf{y};\omega)\\
&+\omega^2(\eps-\eps_0)u_{\mbox{\tiny bac}}^{(l)}(\mathbf{x};\omega)\mathcal{N}(\mathbf{x},\mathbf{y};\omega)\bigg]d\sigma(\mathbf{x}) \frac{\p u_{\mbox{\tiny bac}}^{(j)}(\mathbf{y};\omega)}{\p\boldsymbol{\nu}(\mathbf{y})}dS(\mathbf{y})\\
=&h\int_{\sigma}\bigg[\nabla u_{\mbox{\tiny bac}}^{(l)}(\mathbf{x};\omega)\cdot\mathbb{M}(\sigma;\mathbf{x}) \cdot\nabla u_{\mbox{\tiny bac}}^{(j)}(\mathbf{x};\omega)+\omega^2(\eps-\eps_0)u_{\mbox{\tiny bac}}^{(l)}(\mathbf{x};\omega)u_{\mbox{\tiny bac}}^{(j)}(\mathbf{x};\omega)\bigg]d\sigma(\mathbf{x})\\
=&h\omega^2\int_{\sigma}\bigg((\eps-\eps_0)-\mathbf{d}_l\cdot\mathbb{M}(\sigma;\mathbf{x}) \cdot\mathbf{d}_j^\bot\bigg)e^{-i\omega(\mathbf{d}_j^\bot-\mathbf{d}_l)\cdot\mathbf{x}}d\sigma(\mathbf{x}).
\end{align*}

With this, we can generate a Multi-Static Response (MSR) matrix $\mathbb{A}=(A_{jl}(\mathbf{x};\omega))_{j,l=1}^{L}\in\mathbb{C}^{L\times L}$ whose element $A_{jl}(\mathbf{x};\omega)$ is the collected normalized boundary measurement at observation number $j$ for the incident number $l$:
\[A_{jl}(\mathbf{x};\omega):=\int_{\p\Omega}\bigg(u_{\mbox{\tiny tot}}^{(l)}(\mathbf{y};\omega)-u_{\mbox{\tiny bac}}^{(l)}(\mathbf{y};\omega)\bigg)\frac{\p u_{\mbox{\tiny bac}}^{(j)}(\mathbf{y};\omega)}{\p\boldsymbol{\nu}(\mathbf{y})}dS(\mathbf{y}),\]
for $j,l=1,2,\cdots,L$.

It is worth emphasizing that for a given frequency $\omega=\frac{2\pi}{\lambda}$, based on the resolution limit, any detail less than one-half of the wavelength cannot be retrieved. Hence, if we divide thin inclusion $\Gamma$ into $M$ different segments of size of order $\frac{\lambda}{2}$, only one point, say, $\mathbf{x}_m$, $m=1,2,\cdots,M$, at each segment will affect the imaging (see \cite{A1,A2,AGKPS,AKLP,P1,P2,PL1,PL2,PL3}).

For the sake of simplicity, let us set $\mathbf{d}_j^\bot=-\mathbf{d}_j$, i.e., we have the same incident and observation directions configuration, and assume that $M<L$; then,
\begin{align*}
A_{jl}(\mathbf{x};\omega)=&h\omega^2\int_{\sigma}\bigg((\eps-\eps_0)+\mathbf{d}_l\cdot\mathbb{M}(\sigma;\mathbf{x}) \cdot\mathbf{d}_j\bigg)e^{i\omega(\mathbf{d}_j^\bot+\mathbf{d}_l)\cdot\mathbf{x}}d\sigma(\mathbf{x})\bigg|_{\mathbf{d}_j^\bot=-\mathbf{d}_j}\\
\approx&h\frac{|\sigma|}{M}\sum_{m=1}^{M}\bigg[(\eps-\eps_0)+2\left(\frac{1}{\mu}-\frac{1}{\mu_0}\right) \mathbf{d}_j\cdot\mathbf{t}(\mathbf{x}_m)\mathbf{d}_l\cdot\mathbf{t}(\mathbf{x}_m)\\
&+2\left(\frac{1}{\mu_0}-\frac{\mu}{\mu_0^2}\right)\mathbf{d}_j\cdot\mathbf{n}(\mathbf{x}_m) \mathbf{d}_l\cdot\mathbf{n}(\mathbf{x}_m)\bigg]e^{i\omega(\mathbf{d}_j+\mathbf{d}_l)\cdot\mathbf{x}_m},
\end{align*}
where $|\sigma|$ denotes the length of $\sigma$.

Now, let us perform the Singular Value Decomposition (SVD) of $\mathbb{A}$
\[\mathbb{A}=\mathbb{US}\overline{\mathbb{V}}^T\approx \sum_{m=1}^{M}\mathbf{u}_m(\mathbf{x}_m;\omega)s_m(\omega)\overline{\mathbf{v}}_m^T(\mathbf{x}_m;\omega)\]
and define a vector $\mathbf{w}(\mathbf{x};\omega)\in\mathbb{C}^{L\times1}$ as
\begin{equation}\label{VecW}
  \mathbf{w}(\mathbf{x};\omega)=\bigg(\mathbf{c}\cdot(1,\mathbf{d}_1)e^{i\omega\mathbf{d}_1\cdot\mathbf{x}}, \mathbf{c}\cdot(1,\mathbf{d}_2)e^{i\omega\mathbf{d}_2\cdot\mathbf{x}},\cdots, \mathbf{c}\cdot(1,\mathbf{d}_L)e^{i\omega\mathbf{d}_L\cdot\mathbf{x}}\bigg)^T,
\end{equation}
where the selection of $\mathbf{c}\in\mathbb{R}^3\backslash\set{\mathbf{0}}$ depends on the shape of the supporting curve $\sigma(\mathbf{x})$ (see \cite[Section 4.3.1]{PL3} for a detailed discussion). Then, by defining a projection operator $\mathbb{P}$ onto the null (or noise) subspace, for $L\times L$ identity matrix $\mathbb{I}_L$,
\[\mathbb{P}(\mathbf{w}(\mathbf{x};\omega)):=\left(\mathbb{I}_L -\sum_{m=1}^{M}\mathbf{u}_m(\mathbf{x}_m;\omega)\overline{\mathbf{u}}_m(\mathbf{x}_m;\omega)\right) \mathbf{w}(\mathbf{x};\omega),\]
we can construct MUSIC-type imaging functional:
\begin{equation}\label{MUSIC}
  \mathbb{E}_{\mbox{\tiny MUSIC}}(\mathbf{x};\omega)=\frac{1}{||\mathbb{P}(\mathbf{w}(\mathbf{x};\omega))||}.
\end{equation}

Now, we introduce Kirchhoff migration;
\[\mathbb{E}_{\mbox{\tiny KM}}(\mathbf{x};\omega):=|\overline{\mathbf{w}(\mathbf{x};\omega)}\mathbb{A}\mathbf{w}(\mathbf{x};\omega)| =\sum_{m=1}^{L}s_m(\omega)|\overline{\mathbf{w}(\mathbf{x};\omega)}\mathbf{u}_m(\mathbf{x}_m;\omega)| |\overline{\mathbf{w}(\mathbf{x};\omega)}\overline{\mathbf{v}}_m(\mathbf{x}_m;\omega)|,\]
where $\mathbf{w}(\mathbf{x};\omega)$ is defined in (\ref{VecW}). Note that based on the Statistical Hypothesis Testing, multi-frequency Kirchhoff migration
\begin{align*}
\mathbb{E}_{\mbox{\tiny MKM}}(\mathbf{x};\omega_k):=\sum_{k=1}^{K}\mathbb{E}_{\mbox{\tiny KM}}(\mathbf{x};\omega)=\sum_{k=1}^{K}\sum_{m=1}^{L}s_m(\omega_k)|\overline{\mathbf{w}(\mathbf{x};\omega_k)} \mathbf{u}_m(\mathbf{x}_m;\omega_k)||\overline{\mathbf{w}(\mathbf{x};\omega_k)} \overline{\mathbf{v}}_m(\mathbf{x}_m;\omega_k)|
\end{align*}
will yields a more more accurate result than the single frequency case (see \cite{AGKPS,HSSZ,P1,P2,PL3} for a detailed description).

\section{Proof of Lemma \ref{lem3}}\label{SecB}
Now, we shall show a proof of Lemma \ref{lem3}. For the purpose of simplicity, we set $\eps_0=\mu_0=1$, $\eps>\eps_0$ and $\mu>\mu_0$.

\begin{figure}
\begin{center}
\includegraphics[width=0.5\textwidth,keepaspectratio=true,angle=0]{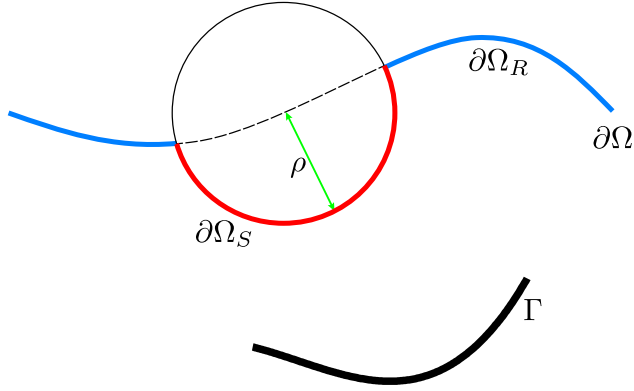}
\caption{\label{FigureSingularIntegral}Illustration of $\p\Omega_S$ (red-colored line) and $\p\Omega_R$ (blue colored line).}
\end{center}
\end{figure}

First, let us explore the structure of $d_T\mathbb{E}_\varepsilon(\mathbf{z};\omega)$ in (\ref{TopologicalDerivative1}). Notice that in this case, $\eps\ne\eps_0$ and $\mu=\mu_0$. Since $v_{\mbox{\tiny adj}}^{(l)}(\mathbf{x};\omega)$ satisfies adjoint problem (\ref{Adjoint1}), it can be represented by the Neumann function $\mathcal{N}(\mathbf{x},\mathbf{y};\omega)$: for $\mathbf{z}\in\Omega$,
\begin{equation}\label{FormulaV}
  v_{\mbox{\tiny adj}}^{(l)}(\mathbf{z};\omega)=\int_{\p\Omega}\frac{v_{\mbox{\tiny adj}}^{(l)}(\mathbf{y};\omega)}{\p\boldsymbol{\nu(\mathbf{y})}}\mathcal{N}(\mathbf{z},\mathbf{y};\omega)dS(\mathbf{y})
=\int_{\p\Omega}\bigg(u_{\mbox{\tiny tot}}^{(l)}(\mathbf{y};\omega)-u_{\mbox{\tiny bac}}^{(l)}(\mathbf{y};\omega)\bigg)\mathcal{N}(\mathbf{z},\mathbf{y};\omega)dS(\mathbf{y})
\end{equation}
Plugging formula (\ref{FormulaV}) into (\ref{TopologicalDerivative1}) and applying asymptotic expansion formula (\ref{AsymptoticExpansionFormula}) yields that
\begin{align}
\begin{aligned}\label{TopEps1}
  d_T\mathbb{E}_\eps(\mathbf{z};\omega)=&\mathfrak{Re}\sum_{l=1}^{L}\bigg(v_{\mbox{\tiny adj}}^{(l)}(\mathbf{z};\omega)\overline{u_{\mbox{\tiny bac}}^{(l)}(\mathbf{z};\omega)}\bigg)\\
  \approx&\mathfrak{Re}\sum_{l=1}^{L}\bigg[\bigg(\int_{\p\Omega}\left(u_{\mbox{\tiny tot}}^{(l)}(\mathbf{y};\omega)-u_{\mbox{\tiny bac}}^{(l)}(\mathbf{y};\omega)\right)\mathcal{N}(\mathbf{z},\mathbf{y};\omega)dS(\mathbf{y})\bigg)\overline{u_{\mbox{\tiny bac}}^{(l)}(\mathbf{z};\omega)}\bigg]\\
  =&h\omega^2(\eps-\eps_0)\mathfrak{Re}\sum_{l=1}^{L}\bigg[\bigg(\int_{\p\Omega}\int_{\sigma}u_{\mbox{\tiny bac}}^{(l)}(\mathbf{x};\omega)\mathcal{N}(\mathbf{x},\mathbf{y};\omega)d\sigma(\mathbf{x})\mathcal{N}(\mathbf{z},\mathbf{y};\omega)dS(\mathbf{y})\bigg)\overline{u_{\mbox{\tiny bac}}^{(l)}(\mathbf{z};\omega)}\bigg]\\
  =&h\omega^2(\eps-\eps_0)\mathfrak{Re}\sum_{l=1}^{L}\int_{\sigma}\bigg(\mathbb{N}(\mathbf{x},\mathbf{z};\omega)u_{\mbox{\tiny bac}}^{(l)}(\mathbf{x};\omega)\overline{u_{\mbox{\tiny bac}}^{(l)}(\mathbf{x};\omega)}\bigg)d\sigma(\mathbf{x})
\end{aligned}
\end{align}
where
\begin{equation}\label{ProductNeumann}
  \mathbb{N}(\mathbf{x},\mathbf{z};\omega):=\int_{\p\Omega}\mathcal{N}(\mathbf{x},\mathbf{y};\omega)\mathcal{N}(\mathbf{z},\mathbf{y};\omega)dS(\mathbf{y}).
\end{equation}
By virtue in \cite{AKL}, Neumann function $\mathcal{N}(\mathbf{x},\mathbf{y};\omega)$ has a logarithmic singularity. Therefore, it can be decomposed into the singular and regular functions;
\begin{equation}
  \mathcal{N}(\mathbf{x},\mathbf{y};\omega)=-\frac{1}{2\pi}\ln|\mathbf{x}-\mathbf{y}|+\mathcal{R}(\mathbf{x},\mathbf{y};\omega),
\end{equation}
where $\mathcal{R}(\mathbf{x},\mathbf{y};\omega)\in C^{1,\alpha}$ in both $\mathbf{x}$ and $\mathbf{y}$ for some $\alpha$ with $0<\alpha<1$ (see \cite{AKL,AKLP} for instance). Since $\mathbf{x}\in\sigma$ and $\mathbf{y}\in\p\Omega$, there is no blow up of $\mathcal{N}(\mathbf{x},\mathbf{y};\omega)$, and it can be bounded by
\[|\mathcal{N}(\mathbf{x},\mathbf{y};\omega)|\leq\frac{1}{2\pi}\ln|\mathbf{x}-\mathbf{y}|+|\mathcal{R}(\mathbf{x},\mathbf{y};\omega)| <\frac{1}{2\pi}\ln\mbox{diam}(\Omega)+\max|\mathcal{R}(\mathbf{x},\mathbf{y};\omega)|,\]
where $\mbox{diam}(\Omega)$ denotes the diameter of $\Omega$. Hence, applying H{\"o}lder's inequality yields
\[|\mathbb{N}(\mathbf{x},\mathbf{z};\omega)|\leq \bigg(\frac{1}{2\pi}\ln\mbox{diam}(\Omega)+\max|\mathcal{R}(\mathbf{x},\mathbf{y};\omega)|\bigg) \int_{\p\Omega}|\mathcal{N}(\mathbf{z},\mathbf{y};\omega)|dS(\mathbf{y}).\]
From the fact that $\mathbf{y}\in\p\Omega$ and $\mathbf{z}\in\Omega$, we must consider the singularity of $\mathcal{N}(\mathbf{z},\mathbf{y};\omega)$ at $\mathbf{z}=\mathbf{y}$ in order to analyze (\ref{ProductNeumann}). For handling this singularity, for a fixed small constant $\rho>0$, generate a ball $B(\mathbf{z},\rho)$ of center $\mathbf{z}$ and radius $\rho$ such that
\[B(\mathbf{z},\rho)\cap\Gamma=\O.\]
Then by separating the boundary $\p\Omega$ into $\p\Omega=\p\Omega_S\cup\p\Omega_R$ (see Figure \ref{FigureSingularIntegral}), where
\[\p\Omega_S=\Omega\cap\p B(\mathbf{z},\rho)\quad\mbox{and}\quad\p\Omega_R=\p\Omega\backslash(\Omega\cap\p B(\mathbf{z},\rho)).\]
Then
\begin{align*}
  \int_{\p\Omega}|\mathcal{N}(\mathbf{z},\mathbf{y};\omega)|dS(\mathbf{y})\leq &\frac{1}{2\pi}\int_{\p\Omega}\ln|\mathbf{z}-\mathbf{y}|dS(\mathbf{y}) +\int_{\p\Omega}|\mathcal{R}(\mathbf{z},\mathbf{y};\omega)|dS(\mathbf{y})\\
  \leq&\frac{1}{2\pi}\lim_{\rho\to0+}\left(\int_{\p\Omega_S}\ln|\mathbf{z}-\mathbf{y}|dS(\mathbf{y}) +\int_{\p\Omega_R}\ln|\mathbf{z}-\mathbf{y}|dS(\mathbf{y})\right)\\
  &+\max|\mathcal{R}(\mathbf{z},\mathbf{y};\omega)|\mbox{length}(\p\Omega)\\
  \leq&\frac{1}{2\pi}\lim_{\rho\to0+}\bigg(\rho\ln\rho+(\mbox{length}(\p\Omega)-\rho)\ln|\mbox{length}(\p\Omega)|\bigg)\\
  &+\max|\mathcal{R}(\mathbf{z},\mathbf{y};\omega)|\mbox{length}(\p\Omega)\\
  =&\mbox{length}(\p\Omega)\bigg(\frac{1}{2\pi}\ln|\mbox{length}(\p\Omega)|+\max|\mathcal{R}(\mathbf{z},\mathbf{y};\omega)|\bigg).
\end{align*}
Here, $\mbox{length}(\p\Omega)$ denotes the length of $\partial\Omega$. Therefore, we can say that $\mathbb{N}(\mathbf{x},\mathbf{z};\omega)$ is bounded by
\begin{align*}
  |\mathbb{N}(\mathbf{x},\mathbf{z};\omega)|\leq&\bigg(\frac{1}{2\pi}\ln\mbox{diam}(\Omega)+\max|\mathcal{R}(\mathbf{x},\mathbf{y};\omega)|\bigg) \times\\
  &\mbox{length}(\p\Omega)\bigg(\frac{1}{2\pi}\ln|\mbox{length}(\p\Omega)|+\max|\mathcal{R}(\mathbf{z},\mathbf{y};\omega)|\bigg)<+\infty,
\end{align*}
and there is no blow up of $\mathbb{N}(\mathbf{x},\mathbf{z};\omega)$.

In this paper, the background solution is selected as $u_{\mbox{\tiny bac}}^{(l)}(\mathbf{x};\omega)=e^{i\omega\mathbf{d}_l\cdot\mathbf{x}}$. Hence, (\ref{TopEps1}) can be written as
\begin{align*}
  d_T\mathbb{E}_\eps(\mathbf{z};\omega)&=h\omega^2(\eps-\eps_0)\mathfrak{Re}\sum_{l=1}^{L}\int_{\sigma}\bigg(\mathbb{N}(\mathbf{x},\mathbf{z};\omega)u_{\mbox{\tiny bac}}^{(l)}(\mathbf{x};\omega)\overline{u_{\mbox{\tiny bac}}^{(l)}(\mathbf{x};\omega)}\bigg)d\sigma(\mathbf{x})\\
  &\sim\mathfrak{Re}\sum_{l=1}^{L}\int_\sigma(\eps-\eps_0)e^{i\omega\mathbf{d}_l\cdot(\mathbf{x-z})}d\sigma(\mathbf{x}).
\end{align*}

Applying the same process to the magnetic permeability contrast case ($\eps=\eps_0$ and $\mu\ne\mu_0$), we can obtain the following structure of $d_T\mathbb{E}_\mu(\mathbf{z};\omega)$:
\[d_T\mathbb{E}_\mu(\mathbf{z};\omega)\sim\mathfrak{Re}\sum_{l=1}^{L}\int_{\sigma}
  \bigg[2\bigg(\frac{1}{\mu}-\frac{1}{\mu_0}\bigg)\mathbf{d}_l\cdot\mathbf{t}(\mathbf{x})
  +2\bigg(\frac{1}{\mu_0}-\frac{\mu}{\mu_0^2}\bigg)\mathbf{d}_l\cdot\mathbf{n}(\mathbf{x})\bigg]e^{i\omega\mathbf{d}_l\cdot(\mathbf{x-z})}d\sigma(\mathbf{x}).\]

\end{document}